\documentclass[iop]{emulateapj}
\usepackage{graphicx}
\usepackage{enumerate}
\usepackage{amssymb, amsmath}
\usepackage{natbib}
\usepackage{color}
\usepackage{ulem}
\usepackage{dblfnote}

\def\pasa {PASA}
\def\procspie{Proc.~SPIE}

\renewcommand{\thefootnote}{\alph{footnote}}

\newcommand{\tokyo}{1}
\newcommand{\abc}{2}
\newcommand{\tokodai}{3}
\newcommand{\hayama}{4}
\newcommand{\chile}{5}
\newcommand{\subaru}{6}
\newcommand{\lablaglange}{7}
\newcommand{\Maxplanck}{8}
\newcommand{\ias}{9}
\newcommand{\charleston}{10}
\newcommand{\munchen}{11}
\newcommand{\gsfc}{12}
\newcommand{\eureka}{13}
\newcommand{\Gca}{14}
\newcommand{\naoj}{15}
\newcommand{\hawaii}{16}
\newcommand{\princeton}{17}
\newcommand{\osaka}{18}
\newcommand{\hiroshima}{19}
\newcommand{\cab}{20}
\newcommand{\jpl}{21}
\newcommand{\sokendai}{22}
\newcommand{\taiwan}{23}
\newcommand{\zurich}{24}
\newcommand{\kavli}{25}
\newcommand{\hokkaido}{26}
\newcommand{\oklahoma}{27}
\newcommand{\tohoku}{28}

\begin{document}

\title{The SEEDS High Contrast Imaging Survey of Exoplanets around Young Stellar Objects}

\author{Taichi Uyama$^\tokyo$}
\author{Jun Hashimoto$^\abc$}
\author{Masayuki Kuzuhara$^\tokodai$}
\author{Satoshi Mayama$^\hayama$}
\author{Eiji Akiyama$^\chile$}
\author{Thayne Currie$^\subaru$}
\author{John Livingston$^\tokyo$}
\author{Tomoyuki Kudo$^\subaru$}
\author{Nobuhiko Kusakabe$^\abc$}
\author{Lyu Abe$^{\lablaglange}$}
\author{Wolfgang Brandner$^{\Maxplanck}$}
\author{Timothy D. Brandt$^{\ias}$}
\author{Joseph C. Carson$^{\charleston,\Maxplanck}$}
\author{Sebastian Egner$^{\subaru}$}
\author{Markus Feldt$^{\Maxplanck}$}
\author{Miwa Goto$^{\munchen}$}
\author{Carol A. Grady$^{\gsfc,\eureka,\Gca}$}
\author{Olivier Guyon$^{\subaru}$}
\author{Yutaka Hayano$^{\subaru}$}
\author{Masahiko Hayashi$^{\naoj}$}
\author{Saeko S. Hayashi$^{\subaru}$}
\author{Thomas Henning$^{\Maxplanck}$}
\author{Klaus W. Hodapp$^{\hawaii}$}
\author{Miki Ishii$^{\naoj}$}
\author{Masanori Iye$^{\naoj}$}
\author{Markus Janson$^{\princeton}$}
\author{Ryo Kandori$^{\naoj}$}
\author{Gillian R. Knapp$^{\princeton}$}
\author{Jungmi Kwon$^\tokyo$}
\author{Taro Matsuo$^{\osaka}$}
\author{Michael W. Mcelwain$^{\gsfc}$}
\author{Shoken Miyama$^{\hiroshima}$}
\author{Jun-Ichi Morino$^{\naoj}$}
\author{Amaya Moro-Martin$^{\princeton,\cab}$}
\author{Tetsuo Nishimura$^\subaru$}
\author{Tae-Soo Pyo$^\subaru$}
\author{Eugene Serabyn$^{\jpl}$}
\author{Takuya Suenaga$^{\naoj,\sokendai}$}
\author{Hiroshi Suto$^{\abc,\naoj}$}
\author{Ryuji Suzuki$^{\naoj}$}
\author{Yasuhiro H. Takahashi$^{\tokyo,\naoj}$} 
\author{Michihiro Takami$^{\taiwan}$}
\author{Naruhisa Takato$^\subaru$}
\author{Hiroshi Terada$^{\naoj}$}
\author{Christian Thalmann$^{\zurich}$}
\author{Edwin L. Turner$^{\princeton,\kavli}$}
\author{Makoto Watanabe$^{\hokkaido}$}
\author{John Wisniewski$^{\oklahoma}$}
\author{Toru Yamada$^{\tohoku}$}
\author{Hideki Takami$^{\naoj}$}
\author{Tomonori Usuda$^{\naoj}$}
\author{Motohide Tamura$^{\tokyo,\abc,\naoj}$}

\footnotetext[\tokyo]{Department of Astronomy, The University of Tokyo, 7-3-1, Hongo, Bunkyo-ku, Tokyo, 113-0033, Japan}
\footnotetext[\abc]{Astrobiology Center of NINS, 2-21-1, Osawa, Mitaka, Tokyo, 181-8588, Japan}
\footnotetext[\tokodai]{Department of Earth and Planetary Sciences, Tokyo Institute of Technology, Ookayama, Meguro-ku, Tokyo 152-8551, Japan}
\footnotetext[\hayama]{The Center for the Promotion of Integrated Sciences, The Graduate University for Advanced Studies (SOKENDAI), Shonan International Village, Hayama-cho, Miura-gun, Kanagawa 240-0193, Japan}
\footnotetext[\chile]{Chile Observatory, National Astronomical Observatory of Japan, 2-21-2, Osawa, Mitaka, Tokyo, 181-8588, Japan}
\footnotetext[\subaru]{Subaru Telescope, National Astronomical Observatory of Japan, 650 North A‘ohoku Place, Hilo, HI96720, USA}
\footnotetext[\lablaglange]{Laboratoire Lagrange (UMR 7293), Universite de Nice-Sophia Antipolis, CNRS, Observatoire de la Coted'azur, 28 avenue Valrose, 06108 Nice Cedex 2, France}
\footnotetext[\Maxplanck]{Max Planck Institute for Astronomy, K$\ddot{o}$nigstuhl 17, 69117 Heidelberg, Germany}
\footnotetext[\ias]{Astrophysics Department, Institute for Advanced Study, Princeton, NJ, USA}
\footnotetext[\charleston]{Department of Physics and Astronomy, College of Charleston, 66 George St., Charleston, SC 29424, USA}
\footnotetext[\munchen]{Universit$\ddot{a}$ts-Sternwarte M$\ddot{u}$nchen, Ludwig-Maximilians-Universit$\ddot{a}$t, Scheinerstr. 1, 81679 M$\ddot{u}$nchen,Germany}
\footnotetext[\gsfc]{Exoplanets and Stellar Astrophysics Laboratory, Code 667, Goddard Space Flight Center, Greenbelt, MD 20771, USA}
\footnotetext[\eureka]{Eureka Scientific, 2452 Delmer, Suite 100, Oakland CA96002, USA}
\footnotetext[\Gca]{Goddard Center for Astrobiology}
\footnotetext[\naoj]{National Astronomical Observatory of Japan, 2-21-1, Osawa, Mitaka, Tokyo, 181-8588, Japan}
\footnotetext[\hawaii]{Institute for Astronomy, University of Hawaii, 640 N. A‘ohoku Place, Hilo, HI 96720, USA}
\footnotetext[\princeton]{Department of Astrophysical Science, Princeton University, Peyton Hall, Ivy Lane, Princeton, NJ08544, USA}
\footnotetext[\osaka]{Department of Earth and Space Science, Graduate School of Science, Osaka University, 1-1 Machikaneyamacho, Toyonaka, Osaka 560-0043, Japan}
\footnotetext[\hiroshima]{Hiroshima University, 1-3-2, Kagamiyama, Higashihiroshima, Hiroshima 739-8511, Japan}
\footnotetext[\cab]{Department of Astrophysics, CAB-CSIC/INTA, 28850 Torrej$\acute{o}$n de Ardoz, Madrid, Spain}
\footnotetext[\jpl]{Jet Propulsion Laboratory, California Institute of Technology, Pasadena, CA, 171-113, USA}
\footnotetext[\sokendai]{Department of Astronomical Science, The Graduate University for Advanced Studies, 2-21-1, Osawa, Mitaka, Tokyo, 181-8588, Japan}
\footnotetext[\taiwan]{Institute of Astronomy and Astrophysics, Academia Sinica, P.O. Box 23-141, Taipei 10617, Taiwan}
\footnotetext[\zurich]{Swiss Federal Institute of Technology (ETH Zurich), Institute for Astronomy, Wolfgang-Pauli-Strasse 27, CH-8093 Zurich, Switzerland}
\footnotetext[\kavli]{Kavli Institute for Physics and Mathematics of the Universe, The University of Tokyo, 5-1-5, Kashiwanoha, Kashiwa, Chiba 277-8568, Japan}
\footnotetext[\hokkaido]{Department of Cosmosciences, Hokkaido University, Kita-ku, Sapporo, Hokkaido 060-0810, Japan}
\footnotetext[\oklahoma]{H. L. Dodge Department of Physics \& Astronomy, University of Oklahoma, 440 W Brooks St Norman, OK 73019, USA}
\footnotetext[\tohoku]{Astronomical Institute, Tohoku University, Aoba-ku, Sendai, Miyagi 980-8578, Japan}

\begin{abstract} 
We present high-contrast observations of 68 young stellar objects (YSOs) explored as part of the SEEDS survey on the Subaru telescope. Our targets are very young ($<$10 Myr) stars, which often harbor protoplanetary disks where planets may be forming. We achieve a typical contrast of $\sim$$10^{-4}$--$10^{-5.5}$ at an angular distance of 1\arcsec\ from the central star, corresponding to typical mass sensitivities (assuming hot-start evolutionary models) of $\sim$10 ${\rm M_J}$ at 70 AU and $\sim$6 ${\rm M_J}$ at 140 AU. We detected a new stellar companion to HIP 79462 and confirmed the substellar objects GQ Lup b and ROXs 42B b. An additional six companion candidates await follow-up observations to check for common proper motion. Our SEEDS YSO observations probe the population of planets and brown dwarfs at the very youngest ages; these may be compared to the results of surveys targeting somewhat older stars. Our sample and the associated observational results will help enable detailed statistical analyses of giant planet formation.

\end{abstract}

\section{Introduction} \label{sec: Introduction}
Since the first convincing detection of an exoplanet around a main sequence star \citep[][]{Mayor1995}, in 1995, about 3500 exoplanets have been confirmed. Surprisingly, many exoplanets, such as hot Jupiters and high-eccentricity planets, are quite different from those in the solar system \citep[e.g.][]{Mayor1995,Holman1997}. 
Those unexpected exoplanets cannot necessarily be reproduced by conventional theories of planet formation, which were originally developed to explain the properties of the solar system \citep[core accretion model; e.g.][]{Hayashi1985}. Such planet formation models \citep[e.g.][]{Ida2013} have been updated to help explain the myriad exoplanets discovered to date. Ongoing observations and characterizations of various exoplanets help  test the updated models.
Most of the planet detections to date  have resulted from radial velocity \citep[e.g.][]{Mayor1995} and transit surveys \citep[e.g.][]{Auvergne2009,Natalie2011}. These indirect methods observe stars and measure periodic fluctuations caused by the existence of orbiting planets. They are particularly effective for detecting exoplanets with shorter orbital periods, like less than 5 years.

Direct imaging is technically difficult due to the large dynamic range required to detect a planet that is $\sim$$10^{-6}$ times fainter than the central star. However, with the development of adaptive optics (AO) on large, ground-based telescopes, the method has  begun to successfully open up the previously unexplored parameter-space occupied by wide-separation, gas giants ($\sim$5--75 ${\rm M_J}$ and $\sim$10--1000 AU). The number of exoplanets confirmed through direct imaging is still far fewer than those found through indirect methods, such as radial velocity or transit. However, direct imaging can reveal physical parameters \citep[e.g. mass, temperature and atmosphere information;][]{Currie2011} and orbital parameters \citep[e.g. semimajor axis, inclination and eccentricity;][]{Chauvin2012} of wide-orbit exoplanets that are poorly explored by indirect methods.
As a result, direct imaging uniquely probes exoplanet populations that are effectively inaccessible by radial velocity and transit methods. The first directly-imaged planets--- e.g. HR 8799 bcde; $\kappa$ And b; $\beta$ Pic b; GJ 504 b \citep[][]{Marois2008,Lagrange2010,Carson2013,Kuzuhara2013}--- prove the extremes of planet formation. These planets, orbiting at $\sim$10-150 AU and with masses of $M\sim$3--15 ${\rm M_J}$, are difficult to form by either core accretion or disk instability \citep[][]{Dodson2009,Kratter2010,Currie2011}. However, constraining the frequency and semimajor axis distribution of these planets may clarify which planet formation mechanism is responsible for their existence \citep[][]{Brandt2014a}.
Estimating physical parameters of detected exoplanets requires accurate age estimations of the host stars. 
Young stellar objects (YSOs) make appealing target populations in part because most of them are associated with star-forming regions whose ages are relatively well established ($\sim$1--10 Myr).

Here we observe YSOs with direct imaging in an effort to detect young exoplanets with ongoing formation and to improve our understanding of how planets form in protoplanetary disks. Previous direct imagin reported a few low-mass companions around YSOs such as GQ Lup b, ROXs 42Bb, and possibly LkCa 15 bc \citep[][]{Neuhauser2005,Currie2014a,Sallum2015}.
Strategic Exploration of Exoplanets and Disks with Subaru \citep[SEEDS;][]{Tamura2009}, a project exploring exoplanets and circumstellar disks with Subaru/HiCIAO \citep[][]{Suzuki2010HiCIAO} and AO188 \citep[][]{Hayano2008}, has conducted, from 2009 to 2015, a direct imaging  survey of more than 400 targets. 
A main goal of SEEDS is to constrain formation and evolution scenarios of planetary systems, including not only exoplanets but also protoplanetary disks, from the point of view of direct imaging. To achieve this, SEEDS selected targets whose ages range from 1 Myr to a few Gyr, for systematic analyses of systems at different stages. Comparing information on exoplanets with various ages can be useful for placing constraints on their formation and evolution mechanisms.

SEEDS has several target categories: Young Stellar Objects, Moving Groups \citep[][]{Brandt2014b}, Open Clusters \citep[][]{Yamamoto2013}, Debris Disks \citep[][]{Janson2013}, and Nearby Stars \citep[][]{Kuzuhara2016}.
We report here the observations and results from systematic explorations carried out in the YSO category. 
The sample selection of YSOs is described in Section \ref{sec: Sample Selection}. In Section \ref{sec: Observations and Data Reduction}, the observations and data reduction are presented for each target. Section \ref{sec: Results} provides detection limit results and expanded information on individual targets. Preliminary statistical discussions are described in Section \ref{sec: Results and Discussion of Total Data}. Finally, we summarize our results in Section \ref{sec: Summary}.

\section{Sample Selection} \label{sec: Sample Selection} 
\subsection{Young Stellar Objects} \label{sec: Young Stellar Objects}

We define YSOs to be young stars  with average ages less than $\sim$10 Myr. Most of our targets show evidence for dusty circumstellar disks, ranging presumably from pre-planet building phases (optically thick, gas rich protoplanetary disks) to post-jovian planet building phases (optically thin, gas poor debris disks).
Some spectral energy distribution (SED) data of YSOs show far-infrared (FIR) excess but little mid-infrared (MIR) excess \citep[][]{Strom1989}. Disks showing such SED features are called  ``transitional disks'' and have been predicted to have gaps, which may be a result of planet formation \citep[][]{Marsh1992,Marsh1993,Quanz2013a}. Some such YSOs with transitional disks exhibit resolved gaps and accompanying structures (e.g. spiral arms) that may indicate hidden planets \citep[][]{Hashimoto2012, Muto2012,Grady2013,Thalmann2014,Currie2014c}. In addition, a few young protoplanets have been imaged inside the gaps of transitional disks \citep[][]{Kraus2012,Currie2015}.
At near infrared (NIR) wavelengths, young planets are brighter  than their older counterparts, so YSOs may be particularly well suited for simultaneously probing lower mass planets and protoplanetary disks (see Section \ref{sec: Observations and Data Reduction} for details).

SEEDS adopts two imaging modes. One  is most effective for disk searches while the other is more optimized for planet searches (see Section \ref{sec: SEEDS Observations}).  The two methods can be combined to simultaneously study both planets and disks.
The simultaneous study of disk and exoplanets is helpful for analyzing the relationship between disk structures and planet formation. 
However, note that our SEEDS explorations are limited to exoplanets outside the gap region, due to self-subtraction effects (details are described in Section \ref{sec: Data Reduction}). Also, we are so far unable to conclude systematic relations between disk geometry and exoplanet frequency.

Section \ref{sec: Results of Individual Companion Survey} provides the available individual SED and disk information for each YSO. 
Most of our YSO targets are located at distances larger than 100 pc. This is because almost all SEEDS/YSO targets belong to star-forming regions (SFRs) and active SFRs in the Milky Way are mainly located in Gould Belt \citep[][]{Dunham2015}, which is farther than 140 pc. The targets of the YSO category are therefore fainter than those of other SEEDS categories, making the AO performance less effective.  Also, the large distances make it more difficult to distinguish orbiting companions at a solar-system scale from the central star.   
However, as previously mentioned, our SEEDS data were optimized for studying disks, but can still place important constraints on giant planets.

\subsection{SEEDS Target Selections}
\label{sec: SEEDS Targets and Selection Criteria}

Table \ref{targets} lists all SEEDS/YSO data analyzed in this work and Figure \ref{splist} summarizes spectral types of the YSO targets. Hereafter the targets are sorted by their RAs.

We explain here the procedures to select our YSO targets. The guidelines and criteria are somewhat different for each SFR (see below for the details). However, we first describe the common target-selection procedures used for all the SFRs. \par
First, we searched the literature nearby SFRs observable from the summit of Mauna Kea.
We aimed to select YSOs  brighter than $R = 15$ to ensure high quality adaptive optics performance and therefore a sufficient angular resolution. We de-prioritized M-type YSOs due to a small frequency of gas giants expected around M dwarfs \citep[][]{Kokubo2002}.
Famous and relatively bright sources that have been observed in various methods such as AB Aur are prioritized.
The YSOs whose disks had been resolved before SEEDS observations are also prioritized.
Our used catalogues are described in the sections for each SFR. Our SEEDS/YSO observations were largely divided into two effective surveys: disk survey and planet survey. 
The YSOs whose SEDs suggested a large, full disk were prioritized for disk study. Objects with transitional or debris disks were selected for both planet and disk explorations. 
Sources exhibiting little or weak IR excesses were slated for planet searches if they showed any youth indicators such as H$\alpha$ emissions, X-ray emissions, or Lithium absorptions. 
For each SFR, we adopted typical estimates for age and distance (see Table \ref{star-forming-groups}), which we later used to constrain potential exoplanet masses  around the YSO. Most of our employed age estimations derived from isochrones of pre-main sequence evolutional tracks \citep[e.g.][]{DM1994,Baraffe1998}. \par
We below describe the basic procedures of target selection for individual SFRs. The majority of targets were selected before the SEEDS survey started, but we include in the subsequent summaries  information published since then. \\
\\
{\it Taurus-Auriga ---} At the first step, we selected YSOs with spectral types of B0--M1 \citep[][]{Strom1989,Beckwith1990,Kenyon1995,Andrews2005,Furlan2006,McCabe2006,Najita2007}. Subsequently, we removed close binaries that have separations between components are less than 3\arcsec\ \citep{Beckwith1990,White2001,White2004,Furlan2006,McCabe2006,Najita2007}. 
The YSOs fainter than $R=15$ were also removed by checking the USNO-B1.0 catalog \citep{Monet2003}. The $R$-band magnitudes of Taurus-Auriga (hereafter Taurus) YSOs measured by our previous Subaru-AO observations \citep[e.g.,][]{Fukagawa2004,Itoh2005} were additionally used to support the USNO-B1.0 photometry.     
Next, we investigated SEDs of all selected YSOs using {\it Spitzer} data \citep[][]{Hartmann2005,Calvet2005,Furlan2006,Padgett2006,Najita2007,Luhman2010,Espaillat2011} and radio observations \citep[][]{Kitamura2002,Andrews2005} in order to find circumstellar disks. We also assign the higher priorities to the YSOs with the observations of resolved disks \citep[][and references in Section \ref{sec: Results of Individual Companion Survey}]{Kitamura2002,Andrews2007a,Isella2010,Karr_2010_CIAO}.
In particular, the transitional disks are the highest-priority targets. A transitinaol disk's inner hole can be caused by grain growth and/or photoevapolation, as well as the influence of planet formation \citep[][]{Williams2011}. 
The large inner hole ($>\sim$100 AU) serves to significantly decrease the infrared continuum \citep[][]{Williams2011}. Hence, objects detected at only longer wavelength (e.g. $>$850 $\mu$m) have high priorities.

We examined CIAO (Subaru) and HST archival data to verify whether or not there are circumstellar structures such as nebular and companions around the targets. 
The stellar jets are closely related with the accretion disks \citep[][]{Blandford1982} and thus we include YSOs which have the jet among disk survey targets.

\cite{Kenyon2008} summarized distance investigations of YSOs in Taurus-Auriga (hereafter Taurus), with the optimum estimate of 140--145 pc as this SFR's distance. On the other hand, distances of Taurus members estimated from their parallaxes are distributed within $\sim$130--160 pc \citep[][]{Torres2007,Torres2009,Torres2012}.
We accordingly assume the Taurus-Auriga SFR's distance to be 140 pc. 
We calculated the median and the standard deviation of Taurus-Auriga's age using the results of \cite{Bertout2007} in which member's ages are estimated with evolutionary tracks of \cite{Siess2000}, resulting in 1.3--13 Myr.
\cite{Kucuk2010} that carried out isochrone analyses with a large sample of YSOs in Taurus and found that the ages of those YSOs are best consistent to 1--3 Myr. Considering these studies, we adopt 1--13 Myr as the typical age of Taurus.\\    
\\
{\it Upper Scorpius (Upper Sco) ---} 
$R$-band magnitudes for the Upper Sco region were compiled from the USNO-B1.0 catalog \citep{Monet2003}.
To identify YSOs, we first investigated targets' H$\alpha$ emission, Lithium absorption, and X-ray emission from \citet{Walter1994}, \citet{Preibisch1998}, and \citet{Kohler2000}, which were also used to obtain their spectral types and $R$ magnitudes.  
To compile the subset suitable for disk and planet explorations, we selected the young targets whose SEDs showed infrared excesses and/or a gap around 24 $\mu$m, as identified via {\it Spitzer} infrared data \citep[][]{Chen2005,Carpenter2006,Carpenter2009,Dahm2009,Dahm2010}. A gap around 24 $\mu$m indicates that the inner part of the disk is cleared \citep[][]{Strom1989,Espaillat2014}.

We assume an age of the Upper Sco SFR to be 9--13 Myr, based on \cite{Pecaut2012} who derived an age using a large sample of YSOs via not only isochrone but also other indicators such as lithium absorptions or H$\alpha$ emissions \citep[][]{Balachandran1990,White2003}. \cite{Zeeuw1999} and \cite{Bertout1999} measured individual distances of YSOs in Upper Sco with Hipparcos data and derived the typical value of 145 pc, which is used in this work. \\
\\
{\it $\rho$ Ophiuchus ---}
We first took stars of $R<15$ in the $\rho$ Ophiuchus area by investigating the literature \citep{Wilking2005, Bouvier1992, Cieza2007, Zacharias2005}. Next, in order to verify the youth of those stars, we attempted to collect the measurements of their H$_{\alpha}$ emissions summarized in \cite{Wilking2005}, \cite{Bouvier1992}, and \cite{Martin1998}. The previous observations of speckle interferometry \citep[][]{Barsony2003,Ratzka2005}, adaptive optics \citep[][]{Cieza2010}, and HST/NICMOS \citep{Allen2002} helped us rule out close binary systems.

We also checked whether the target candidates have the circumstellar disks, because they may have such disks if they are young.  The presences of the infrared excesses in the SEDs and the submillimeter emissions are indicators of circumstellar disk, so we examined if those indicators have been observed for our target candidates, using the {\it Spitzer} data \citep[][]{Cieza2007, Furlan2009, McClure2010, Cieza2010} and submillimeter observations \citep[][]{Andre1994, Andrews2009}.  In selecting disk-bearing YSOs, we prioritized transitional disks, which are useful for both planet searches and disk studies. We used optical spectroscopy \citep[][]{Bouvier1992,Martin1998,Wilking2005} and IR spectroscopy \citep[][]{Luhman1999} to check the strengths of lithium absorption, which is also a youth indicator, and the spectral types of our target candidates. \par
We cite the distance and the age of Ophiuchus from \cite{Lombardi2008} and \cite{Greene1995}.
\cite{Lombardi2008} obtained precise distance of Ophiuchus and Lupus by combining Hipparcos data with extinction maps from Two Micron All Sky Survey (2MASS) data.
\cite{Greene1995} estimated ages of members in Ophiuchus with several evolutionary models. \cite{Greene1995} concluded that the age difference is comparable with observational uncertainties in determining stellar luminosities. \\
\\
{\it Lupus ---} Systematic sample selection in Lupus for SEEDS is not practical because of its low declination. We then include several intriguing objects based on millimeter / submillimeter observations \citep[][]{Hughes1994,Nuernberger1997,Wichmann1999,Joergens2001,Melo2003,Merin2008,Comeron2009}. They have strong emission at 1.3 mm that indicates the presence of a massive disk \citep[e.g.][]{Tsukagoshi2011}.

As mentioned above, \cite{Lombardi2008} estimated distances for Lupus. The estimated typical value of 155 pc is consistent with \cite{Hughes1993}. \cite{Hughes1994} used a large number of stars in Lupus to estimate ages using isochrone. The mode value is 3.2 Myr but the ages range from 0.1 Myr to 10 Myr \citep[][]{Hughes1994}. We adopt 0.1--10 Myr in order to cover this range.\\
\\
{\it Corona Australis ---} We selected bright stars (R $<$ 15 mag) that show the youth indicators \citep[][]{Wilking1997,Chini2003,Forbrich2007}, except close binaries \citep[][]{Kohler2008}. We then evaluate priorities of each YSO and selected transitional disk targets \citep[][]{Hughes2010}.

We adopt the age of Corona Australis (CrA) as 0.1--10 Myr from \cite{Neuhauser2000}.  Previous studies showed CrA's age is between $\sim$1 Myr \citep[][]{Knacke1973} and 6 Myr \citep[][]{Wilking1992}.
\cite{Neuhauser2000} estimated the age of CrA members reported in \cite{Knacke1973} and \cite{Wilking1992}, and the newly detected YSOs, using the evolutionary tracks from \cite{DM1994}.
For the distance, we used 130 pc provided by \cite{Zeeuw1999} and \cite{Bertout1999}, who investigated the distance of CrA in addition to Upper Sco.\\
\\

We also added some YSOs, which are located in isolation or at other regions, to our target list, because they have intriguing features or they are located at nearer distance.
Finally, we note that our target selections were verified by considering the observational efficiency of all SEEDS targets including YSO targets and their observability such as visibility in the allocated nights.

We adopt the age and distance of YSOs if individual observations have been conducted and use these parameters to estimate detection limit in mass unit (see Section \ref{sec: Mass Estimation} for details). For example, distances of YSOs in Upper Sco and CrA have been catalogued in Hipparcos \citep[][]{Zeeuw1999,Bertout1999} respectively.
Note that we assume and refer to the stellar parameters on Table \ref{each-parameter} for discussions of detection limits in Section \ref{sec: Results of Individual Companion Survey} and Section \ref{sec: Mass Estimation}.

\begin{deluxetable*}{lccc}
\tablecaption{Adopted age and distance of target star-forming groups.}
\tablehead{
\colhead{Group}&
\colhead{Age [Myr]}&
\colhead{Distance [pc]} &
\colhead{reference}
}
\startdata
Upper Scorpius (Upper Sco) &9--13&145&\cite{Zeeuw1999}\\&&&\cite{Bertout1999}\\&&&\cite{Pecaut2012} \\ 
Taurus-Auriga &1--13&140&\cite{Bertout2007}\\&&&\cite{Kenyon2008} \\&&&\cite{Kucuk2010}\\
Ophiuchus&0.3--3&120&\cite{Greene1995}\\&&&\cite{Lombardi2008}\\ 
Lupus&0.1--10&155&\cite{Hughes1993}\\&&&\cite{Hughes1994}\\&&&\cite{Lombardi2008} \\ 
Corona Australis (CrA)&0.5--10&130&\cite{Bertout1999}\\&&&\cite{Zeeuw1999}\\&&&\cite{Neuhauser2000}
\enddata
\label{star-forming-groups}
\end{deluxetable*}

\LongTables
\begin{deluxetable*}{llccccc}

\tablewidth{0pt}
\tablecaption{YSO targets in SEEDS observations}
\tablehead{
    \colhead{HD name} &
    \colhead{Other name} &
    \colhead{Group} &
    \colhead{RA\tablenotemark{a}} &
    \colhead{DEC\tablenotemark{a}} &
    \colhead{$R$ mag\tablenotemark{b}} &
    \colhead{Criteria\tablenotemark{c}}
    }
    
\startdata 
\dots&TYC 4496-780-1&isolated& 00 13 40.5677  & +77 02 10.893&9.4&H$\alpha$, FIR excess \\
21997&\dots&Columba& 03 31 53.64694  & -25 36 50.9366&6.3&nearby, Tr-like disk \\
\dots&LkH$\alpha$ 330&Perseus& 03 45 48.28  & +32 24 11.9&11.2&Tr-disk, gap \\
\dots&IRAS 04028+2948 &Taurus-Auriga& 04 05 59.624  & +29 56 38.26 & 12.3 & IR excess \\
\dots&IRAS 04125+2902&Taurus-Auriga& 04 15 42.787  & +29 09 59.77&14.0&Tr-disk \\
\dots&LkCa 4, V1068 Tau&Taurus-Auriga& 04 16 28.109  & +28 07 35.81&11.8&class III, Li \\
\dots&Elias 1, V892 Tau&Taurus-Auriga& 04 18 40.598  & +28 19 15.51 &14.4& circumbinary disk \\
281934&BP Tau&Taurus-Auriga& 04 19 15.83527  & +29 06 26.8927 &11.6& full disk, CO depletion \\
\dots&V819 Tau&Taurus-Auriga& 04 19 26.260  & +28 26 14.30&11.2&class III, FIR excess, Tr-disk \\
283571&RY Tau&Taurus-Auriga& 04 21 57.41003  & +28 26 35.5709&9.7&Tr-like disk, gap \\
284419&T Tau&Taurus-Auriga& 04 21 59.43445  & +19 32 06.4182&9.8&IR excess \\
\dots&LkCa 8, IP Tau&Taurus-Auriga& 04 24 57.080  & +27 11 56.50&12.5 &preTr-disk\\
\dots&DG Tau&Taurus-Auriga& 04 27 04.698  & +26 06 16.31&12.3& disk with jet \\
285846&UX Tau&Taurus-Auriga& 04 30 03.99626  & +18 13 49.4355&9.8&famous, Tr-disk \\
\dots&HL Tau&Taurus-Auriga& 04 31 38.437  & +18 13 57.65&14.2&multi-ring disk\\
\dots&L1551-51, V1075 Tau&Taurus-Auriga& 04 32 09.269  & +17 57 22.75&11.8&class III, Li \\
\dots & GG Tau(A) & Taurus-Auriga & 04 32 30.346 & +17 31 40.64 & 11.5 & circumbinary disk \\
\dots&L1551-55, V1076 Tau&Taurus-Auriga& 04 32 43.732  & +18 02 56.33 &11.6&class III, Li \\
\dots&DL Tau&Taurus-Auriga& 04 33 39.062  & +25 20 38.23&11.9&IR excess \\  
\dots&DM Tau&Taurus-Auriga& 04 33 48.718  & +18 10 09.99&13.4&Tr-disk \\
\dots&CI Tau&Taurus-Auriga& 04 33 52.005  & +22 50 30.18&12.3&full disk, IR excess \\
\dots&DN Tau&Taurus-Auriga& 04 35 27.375  &  +24 14 58.93&11.8&IR excess \\
\dots&LkCa 14, V1115 Tau&Taurus-Auriga& 04 36 19.093  & +25 42 59.08&10.8&class III, Tr-like disk \\
\dots&LkCa 15, V10759 Tau&Taurus-Auriga& 04 39 17.796  & +22 21 03.48&11.6&Tr-disk \\
\dots&LkH$\alpha$ 332/G1, V1000 Tau&Taurus-Auriga& 04 42 07.326  & +25 23 03.23&9.9&H$\alpha$, Tr-like disk \\
\dots&GO Tau&Taurus-Auriga& 04 43 03.095  & +25 20 18.75&14.3&Tr-like disk \\
\dots&GM Aur&Taurus-Auriga& 04 55 10.983  & +30 21 59.54&11.7&Tr-disk \\
282630&LkCa 19&Taurus-Auriga& 04 55 36.961  & +30 17 55.19&10.6&class III, Tr-like disk \\
31293&AB Aur &Taurus-Auriga& 04 55 45.84521  & +30 33 04.2867&7.0 & famous, asymmetric disk\\
282624&SU Aur&Taurus-Auriga& 04 55 59.38527  & +30 34 01.5190&9.2&IR excess, asymmetric disk \\
\dots&V397 Aur&Taurus-Auriga& 04 56 02.022  & +30 21 03.68&11.3&class III, Tr-like disk \\
\dots&V1207 Tau, RX J0458.7+2046&Taurus-Auriga& 04 58 39.74  & +20 46 44.1&11.5&Li, diskless SED \\
31648&MWC 480&Taurus-Auriga& 04 58 46.26534  & +29 50 36.97506&7.8&full disk \\
34282 & \dots& ONC\tablenotemark{d} & 05 16 00.47763  & -09 48 35.4169 & 9.8 & Tr-disk \\
36112&MWC 758&Taurus-Auriga& 05 30 27.52969  & +25 19 57.0823&8.3&Tr-disk \\
36910&CQ Tau&Taurus-Auriga& 05 35 58.46690  & +24 44 54.0950&10.6&Tr-like disk \\
290764&V1247 Ori&ONC\tablenotemark{d}& 05 38 05.2497  & -01 15 21.670&9.9&Tr-disk \\
\dots&TW Hya&TW Hya& 11 01 51.90671  & -34 42 17.0323&10.4&famous, nearby, Tr-disk \\
\dots&PDS 70&Centaurus& 14 08 10.15  & -41 23 52.5&11.3&Tr-disk \\
135344B&SAO 206462&Lupus& 15 15 48.4394  & -37 09 16.026&8.7&Tr-disk, gap \\
139614&\dots&Lupus-Ophiuchus& 15 40 46.3816  & -42 29 53.548&8.2&preTr-disk \\
\dots&GQ Lup&Lupus& 15 49 12.102  & -35 39 05.12&11.0&companion \\
141441&HIP 77545&Upper Sco& 15 49 59.78565  & -25 09 03.559315&9.0&IR excess \\
142315&HIP 77911&Upper Sco&  15 54 41.59596  &  -22 45 58.5086&6.8&debris disk \\
\dots&IM Lup&Lupus& 15 56 09.17658  & -37 56 06.1193&13.4& full disk \\ 
142527&HIP 78092&Lupus& 15 56 41.88986  & -42 19 23.2746&8.3&famous, asymmetric disk, gap \\
\dots&RX J1603.9-2031A&Upper Sco& 16 03 57.677  & -20 31 05.51&11.7 &IR excess \\
\dots&RX J1604.3-2130A, USco J1604&Upper Sco& 16 04 21.66  & -21 30 28.4&11.8&Tr-disk \\
\dots&SZ 91&Lupus& 16 07 11.592  & -39 03 47.54&14.1&Tr-disk \\
144587&HIP 78996&Upper Sco& 16 07 29.92514  & -23 57 02.4498&9.1&debris disk \\
\dots&V1094 Sco, RX J1608.6-3922&Lupus& 16 08 36.183  & -39 23 02.51&12.5&Tr-disk \\
145655&HIP 79462&Upper Sco& 16 12 55.33509  & -23 19 45.7022&9.0&debris disk \\
147137&HIP 80088&Upper Sco& 16 20 50.23268  & -22 35 38.7985&9.0&debris disk \\
\dots&IRAS 16225-2607, V896 Sco&Ophiuchus& 16 25 38.492  & -26 13 54.08&10.8&Tr-like disk \\
\dots & DoAr 25 &Ophiuchus & 16 26 23.678 & -24 43 13.86 & 12.7 & (less-evolved) Tr-disk \\
148040&HIP 80535&Upper Sco& 16 26 29.91023  & -27 41 20.2520&8.3&IR excess \\
\dots&DoAr 28&Ophiuchus& 16 26 47.42  & -23 14 52.2&12.1&Tr-disk \\
\dots&SR 21&Ophiuchus& 16 27 10.278  & -24 19 12.74&13.2&Tr-disk, gap \\
\dots&IRAS 16245-2423, Oph IRS 48&Ophiuchus& 16 27 37.190  & -24 30 35.03&16.7&Tr-disk \\
\dots&ROXs 42B&Ophiuchus& 16 31 15.016  & -24 32 43.70&13.0&companion \\
\dots & DoAr 44, ROX 44 & Ophiuchus & 16 31 33.46 & -24 27 37.3 & 11.7 & preTr-disk \\
\dots&RX J1633.9-2442&Ophiuchus& 16 33 55.61  & -24 42 05.0&14.1&Tr-disk \\
169142&\dots&isolated& 18 24 29.7787  & -29 46 49.371&8.2& gap, companion candidate \\
\dots&MWC 297&Serpens-Aquila& 18 27 39.527  & -03 49 52.05&11.3&Tr-disk \\
\dots&RX J1842.9-3532&CrA& 18 42 57.948  & -35 32 42.69&11.6&Tr-disk \\
\dots&RX J1852.3-3700&CrA& 18 52 17.299  & -37 00 11.95&11.8&Tr-disk \\
179218&\dots&isolated& 19 11 11.25432  & +15 47 15.6388&7.3&IR excess, resolved disk \\
200775&HIP 103763&Cepheus& 21 01 36.91964  & +68 09 47.7639&8.7&flared disk 
\enddata　

\tablenotetext{a}{Values of right ascension and declination taken from \cite{Hog1998}, \cite{Hog2000}, \cite{Cutri2003}, \cite{Zacharias2003}, \cite{Zacharias2005}, and \cite{Van2007}.}
\tablenotetext{b}{$R$-band magnitudes taken from our photometric measurements, the UCAC 4 \citep[][]{Zacharias2012}, and the USNO-B1.0 catalog \citep[][]{Monet2003}.}
\tablenotetext{c}{Motivations of target selections, whose details and references are described in Section \ref{sec: Results of Individual Companion Survey}.}
\tablenotetext{d}{Orion Nebula Cluster}

\label{targets}
\end{deluxetable*}

\section{Observations and Data Reduction} \label{sec: Observations and Data Reduction}

SEEDS observed about 70 YSOs (more than 100 data sets in total) from 2009 October to 2015 December. 
SEEDS has adopted polarization differential imaging \citep[PDI;][]{Kuhn2001} technique together with angular differential imaging \citep[ADI;][]{Marois2006} in order to efficiently observe both exoplanets and circumstellar disks around those YSO targets. The PDI uses polarization and is useful to image circumstellar disks, since the scattered light from disk surfaces tends to be polarized. Using PDI, SEEDS has mainly reported the results of disk observations for YSOs. However, this technique is not suitable to the exoplanet detections, because self-luminous exoplanets at wide orbits are basically unpolarized. Meanwhile, the ADI takes advantages of field rotation to remove stellar halo and speckles. This method is the most sensitive way to find such a self-luminous exoplanet, since it does not make use of polarization, and can be applied to a detection of point source. 
Our YSO observations were conducted using the ADI and PDI techniques simultaneously to search for planets and analyze disks with the identical data set. Thus, ordinary and extra-ordinary rays are simultaneously obtained by dividing one frame into two or four sub-frames. However, note that our data reductions detailed in Section \ref{sec: ADI Data Reduction} combine all sub-frames in each frame, so the polarization information is not used in the PSF subtractions and only the ADI PSF subtractions are applied to data reductions. We here explain the observations and data reductions to search for exoplanet around the SEEDS/YSO targets.

\subsection{SEEDS Observations} \label{sec: SEEDS Observations}

Table \ref{observations} lists employed filters, imaging modes and observation dates for each target. As shown in Table \ref{observations}, we have observed some targets at multiple nights, for following up detected companion candidates (CCs), characterizing disks, and compensating for previous poor-quality observations.  In order to test whether the CCs are physically associated with their primary stars, we need to confirm that the CCs share the common proper motions (CPMs) with the primary stars.  \par 
Our direct imaging observations were conducted using the adaptive optics system AO188 \citep{Hayano2008} on the Subaru Telescope. All but 3 YSOs have been observed using them as the natural guide stars (NGSs) for AO188. However, IRS 04125+2902, IRAS 16245-2423 and RX J1633.9-2422 are too faint to be observed with the NGS mode, and we accordingly chose the laser guide star mode for those faint targets. \par 
The near-infrared camera HiCIAO \citep{Suzuki2010HiCIAO} was used simultaneously with AO188. Then, multiple differential-imaging modes are available; we can use the `standard PDI' (sPDI) and `quad PDI' (qPDI) modes besides the standard ADI mode. We explain all of modes below.

\begin{figure}
    \centering
    \includegraphics[scale=0.25]{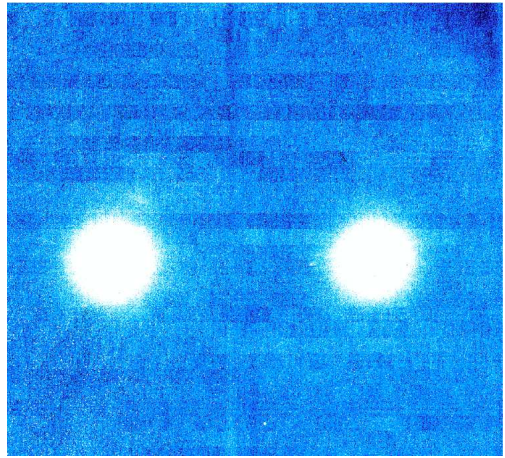}
    \caption{A Subaru/HiCIAO raw image taken in sPDI+ADI mode.}
    \label{sPDI image}
\end{figure}

\begin{figure}
    \centering
    \includegraphics[scale=0.25]{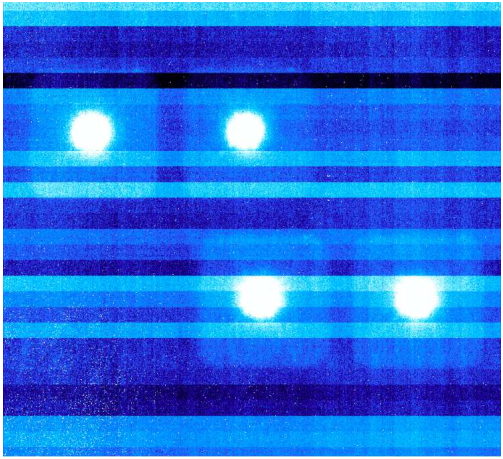}
    \caption{A Subaru/HiCIAO raw image taken in qPDI+ADI mode.}
    \label{qPDI image}
\end{figure}

Subaru/HiCIAO adopts sPDI where we can get ordinary and extra-ordinary rays simultaneously with one Wollaston prism and qPDI where we can get two ordinary and extra-ordinary rays simultaneously with 2 Wollaston prisms. Therefore sPDI show 2 images in one frame and qPDI show 4 images in one frame (see Figure \ref{sPDI image} and \ref{qPDI image}).
HiCIAO PDI observations set an angular offset to decrease instrumental effects of polarization.

The field of views (FOVs) are $\sim$20\arcsec\ $\times$ $\sim$20\arcsec, $\sim$20\arcsec\ $\times$ $\sim$10\arcsec\ (or $\sim$10\arcsec\ $\times$ $\sim$20\arcsec\ before 2014 April), and $\sim$5\arcsec\ $\times$ $\sim$5\arcsec\ for the ADI, sPDI, and qPDI modes, respectively  (see Figure \ref{sPDI image} and \ref{qPDI image}). We found that the new FoV configuration of sPDI can improve the PSF subtractions. 
Our observations were basically performed using an $H$-band ($\sim$1.6 $\mu$m) filter, but a $K_{\rm{s}}$-band ($\sim$2.2 $\mu$m) filter was used in the case of follow-up observations or bad-seeing conditions. \par 

We explore a faint companion around a YSO by acquiring deep long-integration images. 
While a part of our YSO targets is faint enough to be observable without saturating their images, the point spread functions (PFSs) of bright targets were weakly saturated to increase the integration times. 
In addition, we observed some YSOs using occulting masks, helping us avoid excessive saturation and increase integration times. Table \ref{observations} lists the targets observed with occulting masks and the mask sizes.  \par
In addition to the saturated frames, we took unsaturated and unmasked frames for a YSO target using neutral density (ND) filters with transmittance of 9.74\% at $H$ (10.5\% at $K_{\rm{s}}$), 0.854\% at $H$ (1.14\% at $K_{\rm{s}}$), or 0.063\% at $H$ (0.138\% at $K_{\rm{s}}$). These are used to verify the centroid measurements for the masked or saturated images of central star (see Section \ref{sec: Data Reduction}).  Also, these frames are used to measure the contrast between a detected companion and the primary star. \par     
The images of globular cluster M5 or M15 were obtained in the HiCIAO runs, and are compared with archival data from {\it Hubble Space Telescope} to measure the field distortion, including the plate scale and offset angle between the detector's vertical axis and the celestial north direction.    

We measured the HiCIAO's plate scales, which depend on the optical configurations \citep[][]{Brandt2013} and range from $\sim$9.46 to $\sim$9.68 mas pixel$^{-1}$ along X axis and $\sim$9.77 to $\sim$10.02 mas pixel$^{-1}$ along Y axis. 
While the pixel scales vary along the X- and Y-axes, the aspect ratio has been extremely stable. 
Furthermore, the orientation angles between the HiCIAO's vertical axis and celestial north have been stably fixed to $\sim$0\fdg3--0\fdg2. 
These measurements are proper to the HiCIAO's (ADI+)sPDI or qPDI modes, though the almost same results have been obtained by analyzing the globular cluster data taken in the ADI mode alone \citep[][]{Brandt2013}.  
After the distortion correction, the plate scale is corrected to be 9.5 mas pixel$^{-1}$ and the orientation angle is corrected to be zero. 

\begin{deluxetable*}{llccccc}
\tablewidth{0pt}

\tablecaption{SEEDS/YSO Observing logs}
\tablehead{
\colhead{HD name} &
\colhead{Other name} &
\colhead{$t_{\rm tot}$} &
\colhead{Rotation angle} &
\colhead{Band} &
\colhead{Mode} &
\colhead{Date (HST)} \\
\colhead{} &
\colhead{} &
\colhead{[min]} &
\colhead{[deg]} &
\colhead{} &
\colhead{} &
\colhead{}
}

\startdata

\dots&TYC 4496-780-1&54&41.7& $H$ &qPDI+ADI &2012 Jul 10\\
21997&\dots&36&13.8 & $H$ &sPDI+ADI+0\farcs4mask&2013 Jan 3   \\
\dots&LkH$\alpha$ 330&27&10.2& $H$ &sPDI+ADI+0\farcs4mask&2011 Dec 22  \\
&&22&37.3& $H$ &qPDI+ADI&2014 Oct 9 \\
&&24&52.1& $K_{\rm s}$&qPDI+ADI&2015 Jan 11 \\
\dots&IRAS 04028+2948&45&31.5 & $H$ &qPDI+ADI&2012 Dec 31 \\
\dots&IRAS 04125+2902&24&15.4 & $H$ &sPDI+ADI&2013 Nov 24 \\
&&17&9.1& $H$ &sPDI+ADI&2014 Jan 20 \\
\dots&LkCa 4, V1068 Tau&23& 26.1 & $H$&qPDI+ADI&2012 Nov 3 \\
\dots&Elias 1, V892 Tau&24&5.3 & $H$ &sPDI+ADI+0\farcs4mask&2013 Feb 26   \\
281934&BP Tau&32&15.5 & $H$ &qPDI+ADI&2012 Sep 13   \\
\dots&V819 Tau&33&12.8 & $H$ &qPDI+ADI&2012 Jan 1  \\
283571&RY Tau&6&89.2& $H$ &sPDI+ADI+0\farcs4mask&2011 Jan 27 \\
&&19& 77.2 & $H$ &sPDI+ADI&2014 Oct 6  \\
284419&T Tau&12&1.6 & $H$ &qPDI+ADI&2015 Jan 8   \\
\dots&LkCa 8, IP Tau&31&29.2& $H$ &qPDI+ADI&2011 Dec 30   \\
\dots&DG Tau&26 &6.1& $H$ &qPDI+ADI&2015 Jan 11 \\
285846&UX Tau&30&8.9& $H$ &qPDI+ADI&2013 Nov 24  \\
\dots&HL Tau&40&3.6 & $K_{\rm s}$ &qPDI+ADI&2015 Jan 8 \\
\dots&L1551-51, V1075 Tau&26&1.8 & $H$ &qPDI+ADI&2012 Nov 4  \\
\dots&GG Tau(A)&33&0.9& $H$ &sPDI+ADI+0\farcs6mask&2011 Sep 4   \\
\dots&L1551-55, V1076 Tau&26&2.5 & $H$ &qPDI+ADI& 2012 Nov 5 \\
\dots&DL Tau&30&72.3 & $K_{\rm s}$ &qPDI+ADI&2014 Oct 6  \\
\dots&DM Tau&12&162.1& $H$ &sPDI+ADI&2009 Dec 23 \\
\dots&CI Tau&32&130.4 & $H$ &qPDI+ADI&2012 Sep 13  \\
\dots&DN Tau& 12 & 161.4 & $H$ &sPDI+ADI&2009 Dec 22 \\
\dots&LkCa 14, V1115 Tau&28&27.0& $H$ &sPDI+ADI&2011 Dec 29  \\
\dots&LkCa 15, V1079 Tau&11& 134.8 & $H$ &sPDI+ADI&2009 Dec 25  \\
&&11&10.6& $H$ &ADI&2010 Jan 22 \\
&&40&4.5& $K_{\rm s}$ &sPDI+ADI&2010 Dec 1 \\
&&5&11.5& $K_{\rm s}$ &sPDI+ADI&2011 Jan 27  \\
&&34&4.4& $H$ &qPDI+ADI&2013 Nov 12   \\
\dots&LkH$\alpha$ 332/G1, V1000 Tau&28&4.0 & $H$ &sPDI+ADI+0\farcs4mask&2013 Jan 3    \\
\dots&GO Tau&40&16.7& $H$ &sPDI+ADI&2009 Dec 24 \\
\dots&GM Aur&12&31.8& $H$ &qPDI+ADI&2010 Dec 2　\\
&&22&100.7& $H$ &sPDI+ADI&2011 Dec 22  \\
282630&LkCa 19&31&51.1 & $H$ &sPDI+ADI&2011 Dec 23  \\
31293&AB Aur& 8 & 50.1 & $K_{\rm s}$ &sPDI+ADI&2009 Dec 24 \\
282624&SU Aur& 24&14.3& $H$ &qPDI+ADI&2014 Jan 19  \\
&&13&47.9& $K_{\rm s}$ &qPDI+ADI&2014 Oct 9  \\
\dots&V397 Aur&23&75.2& $H$ &sPDI+ADI&2011 Dec 31   \\
\dots&V1207 Tau, RX J0458.7+2046&17&125.0 & $H$ &qPDI+ADI&2012 Nov 3  \\
&&20&57.6& $H$ &sPDI+ADI&2015 Dec 28\\
31648&MWC 480&52&69.9& $H$ &sPDI+ADI&2010 Jan 24 \\
34282&\dots&38&11.5& $H$ &sPDI+ADI+0\farcs4mask&2011 Dec 29   \\
36112&MWC 758&25&6.9& $H$ &sPDI+ADI+0\farcs4mask&2011 Dec 23　\\
&&17&14.3& $K_{\rm s}$ &qPDI+ADI&2011 Dec 25 \\
&&30&102.1& $H$ &qPDI+ADI&2013 Oct 16 \\
36910&CQ Tau&30&2.1& $H$ &sPDI+ADI&2012 Jan 1 \\
&&28&4.0& $H$ &sPDI+ADI+0\farcs3mask&2014 Oct 11\\
&&9&3.2& $H$ &sPDI+ADI&2015 Dec 30 \\
290764&V1247 Ori&42&31.7 & $H$ &sPDI+ADI+0\farcs3mask &2013 Nov 23　\\
&& 36 & 29.9 & $H$ & qPDI+ADI &2014 Jan 19\\
&& 30 & 35.2 & $K_{\rm s}$ & ADI & 2015 Jan 7 \\
\dots&TW Hya&23&13.3& $H$ &qPDI+ADI&2011 Mar 25\\ 
&&30&12.2& $H$ &qPDI+ADI&2013 Jan 3  \\
\dots&PDS 70&23&24.1& $H$ &qPDI+ADI&2012 Feb 27  \\
135344B&SAO 206462&18&23.4 & $H$ &sPDI+ADI+0\farcs4mask&2011 May 20  \\
139614&\dots&26&15.2 & $H$ &sPDI+ADI+0\farcs4mask&2014 Jun 6   \\
\dots&GQ Lup&30&16.1 & $H$ &qPDI+ADI&2013 May 17 \\
&&48&14.3& $H$ &sPDI+ADI+0\farcs4mask&2013 May 19 \\
141441&HIP 77545&31&34.6 & $H$ &sPDI+ADI&2012 May 12 \\
&&25&14.7& $H$ &ADI&2014 Jun 7  \\
142315&HIP 77911& 32 & 28.2 & $H$ & sPDI+ADI & 2012 May 10 \\
&& 20 & 10.5 & $H$ & ADI & 2014 Jun 8   \\
\dots&IM Lup&28&17.7 & $H$ &qPDI+ADI&2014Jun 7   \\
142527&HIP 78092&14&19.3 & $K_{\rm s}$ &sPDI+ADI+0\farcs4mask&2011 May 25 \\
&&12&20.4& $H$ &sPDI+ADI&2012 Apr 10  \\
\dots&RX J1603.9-2031A&52&22.9& $H$ &qPDI+ADI&2012 Jul 9 \\
\dots&RX J1604.3-2130A, USco J1604&25&10.3& $H$ &qPDI+ADI&2012 Apr 11 \\
\dots&SZ 91&44&42.7& $K_{\rm s}$ &sPDI+ADI&2012 May 10  \\
144587&HIP 78996& 34&29.8& $H$ &sPDI+ADI+0\farcs4mask&2011 May 24 \\
&&10&3.9& $H$ &sPDI+ADI&2013 May 17 \\
\dots&V1094 Sco, RX J1608.6-3922&36&21.2 & $H$ &sPDI+ADI+0\farcs4mask&2011 May 20  \\
145655&HIP 79462&40&12.8& $H$ &ADI&2012 Apr 11 \\
&&27&14.8& $H$ &ADI&2014 Apr 22  \\
&&32&25.3& $H$ &ADI&2015 Apr 29 \\
147137&HIP 80088&10&5.7 & $H$ &ADI&2012 Jul 5 \\
&&37&16.6& $H$ &sPDI+ADI&2012 Jul 10  \\
\dots&IRAS 16225-2607, V896 Sco&29&32.0& $H$ &qPDI+ADI&2012 May 11   \\
\dots&DoAr 25&33&25.3 & $H$ &qPDI+ADI&2012 May 15 \\
&&11&7.2& $H$ &ADI&2014 Jun 9 \\
148040&HIP 80535&40&36.6& $H$ &sPDI+ADI&2012 May 13 \\
&& 10&8.5& $H$ &ADI&2015 Apr 29\\
\dots&DoAr 28&48&30.5& $H$ &qPDI+ADI&2012 Jul 8 \\
&&32&22.2& $H$ &qPDI+ADI&2014 Jun 8\\
\dots&SR 21&18&10.7 & $H$ &qPDI+ADI&2011 May 22   \\
\dots&IRAS 16245-2423, Oph IRS 48&28&14.3 & $H$ &qPDI+ADI&2013 May 19  \\
\dots&ROXs 42B&15&11.7 & $Y$ &ADI&2014 Jun 8  \\
\dots&DoAr 44, ROX 44&30&26.7& $H$ &qPDI+ADI&2011 May 22 \\
&&6&4.4& $K_{\rm s}$ &sPDI+ADI&2012 May 12  \\
\dots&RX J1633.9-2442&46&22.8& $H$ &qPDI+ADI&2014 Apr 23 \\
169142&\dots&40&23.1 & $H$ &sPDI+ADI+0\farcs4mask&2011 May 23 \\
&&30&13.2& $J$ &sPDI+ADI+0\farcs4mask&2013 May 19   \\
\dots&MWC 297&20&36.2 & $H$ &sPDI+ADI&2012 Jul 7  \\
\dots&RX J1842.9-3532&34&13.8 & $H$ &qPDI+ADI&2012 Sep 13   \\
\dots&RX J1852.3-3700&33&29.2& $H$ &sPDI+ADI&2011 Sep 5  \\
179218&\dots&34&2.3 & $H$ &sPDI+ADI+0\farcs4mask&2012 Sep 12 \\
200775&HIP 103763&28&26.5& $H$ &sPDI+ADI+0\farcs4mask&2011 Sep 4 
\enddata
\label{observations}
\end{deluxetable*}

\subsection{Data Reduction} \label{sec: Data Reduction}
\subsubsection{ADI Data Reductions}
\label{sec: ADI Data Reduction}
Our YSO targets have been observed using PDI+ADI mode. The data reduction pipelines previously used to search for exoplanets \citep[e.g.,][]{Kuzuhara2013, Brandt2013} are dedicated to the data obtained with only ADI technique. To reduce the ADI+PDI data, we have customized an IDL pipeline for ADI reductions, and developed some auxiliary routines that are based on Python and PyRAF. \par 

For sPDI+ADI or qPDI+ADI data, we first remove their correlated read noise (i.e., destriping), correct for hot pixels, and perform flat-fielding, similarly as for standard SEEDS ADI data \citep[e.g.,][]{Brandt2013}. After these processes, each frame is divided into 2 and 4 sub-frames in the case of sPDI and qPDI observations, respectively. In the sPDI+ADI mode, each sub-frame contains either ordinary ray or extra-ordinary ray for a YSO target. Meanwhile, in the qPDI+ADI mode, the four sub-frames contains two ordinary and extra-ordinary rays. The centroids of target on each sub-frame are individually estimated and each image for the target is shifted to a common center, after correcting field distortion of all sub-frames; the distortion-corrected plate scale is 9.5 mas pixel$^{-1}$. We subsequently integrate the sub-images that were simultaneously acquired into a single image. Finally, the ADI reductions are applied to the sequence of images made by integrating each simultaneous image.
When measuring a detected object's celestial coordinate relative to its central star, we corrected the artificial angular offset in our PDI observations (see Section \ref{sec: SEEDS Observations}).

In order to estimate the stellar centroids, we take into account the three different methods explained below. The first method is applied to a target observed with the long sequence of unsaturated data, for which we calculate the center of a PSF by fitting a 2-dimension elliptical Gaussian function to the PSF on each image with IRAF. The second method is applied to saturated data. Then, we choose a reference frame from all saturated frames for a target, and calculate the relative centroids by comparing the reference frame with the rest of saturated frames.  As in Section \ref{sec: SEEDS Observations}, we obtained unsaturated frames for a YSO target, in the same night. The unsaturated frames are used to verify the centroid of reference frame. \par

For masked data, we adopt the third method, which estimates a position of star by fitting a Moffat function to its masked PSF. In the fit, we exclude the masked area of PSF. Moffat function is given as:                  

\begin{eqnarray}
I(r) = I_0\Bigl[1+\Bigl(2\frac{\sqrt{ (x-x_c)^2 + (y-y_c)^2}}{W} \Bigr)^2\Bigr]^\beta, 
\label{moffat function}
\end{eqnarray}
where $I_0$ is the peak value of PSF, $x_c$ and $y_c$ are the $x$ and $y$ center of PSF, $W$ is FWHM, and $\beta$ is atmosphere scattering coefficient \citep[][]{Moffat1969}. Using least-square fit, each parameter in Equation (\ref{moffat function}), including $x_c$ and $y_c$, is determined. This method based on Moffat function fit should be the best way to determine a center of image whose PSF core is largely masked \citep[0\farcs3--0\farcs4 in our observations;][]{Walker1994}. \par

The AO188 keeps target's centroids stable during observations. Furthermore, atmospheric dispersion corrector \citep[ADC;][]{Egner+10} in AO188 is available to help mitigate the drift of centroids.  Typically, centroid drifts are less than 1--2 pixels ($\sim$10--20 mas) over long sequence of integrations for each target \citep{Brandt2013}. Even in observations at high airmass \citep{Thalmann_2011_HR4796A}, the centroid drift is no more than a few tens of mas \citep{Brandt2013}.  The image-registration algorithms explained above can remove the remaining centroid drifts that are not corrected by AO188; a drift is typically found to be less than $\sim$10 mas. \par

In order to improve the sensitivities of our observations, it is crucial to acquire as long integrations as possible. However, a central star's PSF is usually made more saturated as the integration time of an individual frame increases. We note the balance between a PSF saturation and an integration time.
After subtracting a radial profile from each intensity image, we perform ADI-LOCI data processing \citep{Lafreniere2007} to attenuate starlight and speckles. LOCI requires to tune some parameters, which affect the performance of data reductions. At a radial distance ($=R$) from each central star, our LOCI processing requires a minimum field rotation of 0.75 ($= N_{\rm{\delta}}$) $\times$ PSF's FWHM $/\ R$ radian between a PSF-subtracted image and the PSF-reference images. Our software masks the regions that do not have field rotations large enough to pass this criterion.  In addition, an optimization zone for reference-PSF modeling has an area of 300 ($=N_{\rm{A}}$) $\times$ $\pi (\rm{FWHM}/2)^2$ pixels. See \citet{Lafreniere2007} for details of LOCI parameters. All the LOCI-processed frames are median-combined into a final high-SN image. \par

LOCI partially decreases a flux of point source due to self-subtractions of its PSFs. This flux loss is more significant at smaller separations \citep[][]{Lafreniere2007}. We estimate the artificial flux loss applying LOCI to the data with injected fake companions. We initially inject the fake sources with SN ratios of $\sim$15, from $\sim$0\farcs2 to the edge of the FOV. 
This procedure is repeated 15 times changing initial position angle to cover the full range of ($r,\theta$) in the field avoiding the overlap of PSFs. Finally, we measure the flux loss at each position of more than 1000 fake companions and plot the radial profile of self-subtraction. As to SEEDS data, the typical flux loss is $\sim$50\% at 0\farcs3, $\sim$20\% at 0\farcs8, and under 10\% at larger distances than 1\arcsec.
In addition, we correct the flux loss to derive the photometry of detected companion and contrast limit on each target (see Section \ref{sec: Detection Limit}).

\subsubsection{Producing Contrast Curves}
\label{sec: Detection Limit}
Discussion about exoplanet frequency based on statistical analysis requires detection limits for each target. We calculate the detection limit by defining a noise distribution. First, the final reduced image for each target is normalized by dividing the image by its integration time. Next, the normalized map is convolved with an aperture, whose radius is equal to half of the central star PSF's FWHM. An aperture of the same size was used for the photometry of the central star in unsaturated frames. 
Finally, we define the rings at intervals of $\sim$0\farcs06\ from the central stars and measure the standard deviation of counts at each ring in the convolved map. 
The standard deviations are determined to be the noises as a function of angular separations from the central star.  
To produce a contrast curve, the noise function is divided by the flux of the central star's unsaturated PSFs. 
Some SEEDS/YSO observations in the early part took reference frames with occulting masks or did not take unsaturated frames. These observations have disadvantages in that we have to use other observations for photometric reference, which increases the uncertainty of the contrasts. 
The partial flux losses in the contrast curves have been corrected. 
Finally, we adopt 5$\sigma$ detection threshold in this paper. 

Sensitivity and contrast depend on weather condition, AO performance (seeing), and integration time. Some data were taken in bad condition whose detection limit was not good and are not meaningul for the exoplanet survey due to low contrast. We note that accurately estimating the flux loss by self-subtraction at small angular separations is a difficult task.

\section{Results}
\label{sec: Results}

Among our whole observations, a few observations were conducted with only PDI that cannot be used for exoplanet search. Excluding these, we finally reduced the data taken for the 68 YSO targets (99 data sets in total) to explore planetary-mass compamion candidate.
Note that some data have the rotation angles that are too small to explore inner ($<$ 100 AU) region.  We do not include sources with signal-to-noise ratios (SNRs) less than 5 in the list of companion candidates. 
As a result, we found 15 new companion candidates within 400 AU from their central stars. 
We detected a new stellar companion physically associated with HIP 79462 (see Section \ref{sec: HIP 79462}). We also detected an unreported bright source around TYC 4496-780-1 (see Section \ref{sec: TYC 4496-780-1}) which is either a stellar companion or a background star.
In addition, we confirmed 2 convincing low-mass companions. These companions have been reported previously; GQ Lup b (see Section \ref{sec: GQ Lup}) and ROXs 42B b (see Section \ref{sec: ROXs 42B}). \par
Although our observations detected more point sources,  it is impossible to statistically explore the objects at the projected separations larger than 400 AU because a lot of targets have been observed with qPDI, which has a FOV corresponding to 400 AU. 
Moreover, we assume that planet formations at such large separations from primary stars are relatively challenging since the standard disks around pre-main sequence star should not be so large \citep[c.f.][]{Andrews2007a}, less prioritizing the follow-up observations of those wise-separation companions.  Note that there are the existences of substellar companions at the very large distances from their primary stars \citep[e.g.][]{Bailey14_HD106906,Naud2014}. With the caveat that such substellar companions with very wide separations cannot be ruled out as the companions formed from the circumstellar disks, we do not discuss the companion candidates at the separations larger than 400 AU, based on the above reasons.
The 7 point sources out of 15 candidates within 400 AU from their central stars are identified to be background stars by conducting common proper motion test.

We present the detailed frameworks of detection limits and luminosity-mass conversion in Section \ref{sec: Contrast} and \ref{sec: Mass Estimation}. We also summarize our observed results for individual targets.

\subsection{Contrast}
\label{sec: Contrast}
Our results of detection limit are listed in Table \ref{all contrast}. 
Note that SEEDS observations are carried out basically in $H$-band, but some YSOs are observed in only $K_{\rm s}$-band. For the YSOs observed at multiple epochs, we adopt the deepest detection limits among all detection limits from multiple observations.
Figures \ref{contrast_all_1} and \ref{contrast_all_2} show our final results of 5$\sigma$ detection limit with bright ($<$8 mag in $H$ or $K_{\rm s}$-band) and faint ($\geq$8 mag) stars. Sources brighter than contrast curves can be detected by HiCIAO observations with more than 5$\sigma$ significance. 
Comparing Figures \ref{contrast_all_1} with \ref{contrast_all_2}, we find that typical detection limits are better in observing bright YSOs by a factor of $\sim$2--10 than faint YSOs. We calculated the typical limiting magnitudes by adding the brightness of central stars to the contrast curves, in order to investigate whether or not this difference arises from the difference of AO efficiency. As a result, the limiting magnitudes around bright and faint targets are almost the same; $\sim$14--15 mag at 0\farcs3, $\sim$15.5--17 mag at 0\farcs5, $\sim$18--19 mag at 1\farcs0, $\sim$19--20 mag beyond 2\farcs0.

As already described in Section \ref{sec: Detection Limit}, some of early observations did not take unsaturated frames, thus we need to use the data of the other targets as photometric reference, making it more uncertain to estimate the contrast limits. Actually, a bundle of contrast curves in Figure \ref{contrast_all_1} and \ref{contrast_all_2} look wider than those by Moving Group targets \citep[][]{Brandt2014b,Brandt2014a} to some extent. In CQ Tau, GQ Lup, HIP 103763, HIP 79462, IRAS 16225-2607, MWC 297, PDS 70, ROX44, TYC 4496-780-1 and UX Tau data, there are bright point sources affecting contrast curves. We masked these bright sources when deriving detection limits but we cannot completely remove their influences on detection. IRAS 04125+2902 and LkHa 332 G1 also have bright sources in the FOV, but the companions are outside of our explored separation range. We typically achieve contrast of $\sim$$10^{-3.5}$ at 0\farcs5, $10^{-4}$--$10^{-5}$ at 1\arcsec and $10^{-4.5}$--$10^{-6}$ beyond 2\arcsec. 
This detection limit is similar to that of other SEEDS categories \citep[][]{Brandt2014b} and one of the highest contrasts among the YSO imaging surveys conducted so far (see Section \ref{sec: YSO Surveys}). 
There are a lot of bad pixels at the edges of detector which can effect the contrast curves, and some contrast curves actually become shallower at the wider separations.

Figure \ref{comparison} compares the detection limits of this work with some of previous high contrast surveys. Note that some previous works set detection significances different from 5$\sigma$ and adopted different filters in imaging observations. In comparison with the results of other high contrast imaging surveys for young moving groups (see Section \ref{sec: Young Moving Group Surveys}), our median contrast limits appear to be lower at 1\arcsec. YSOs are basically located at farther distances and thus fainter than young moving group members, affecting AO performance. Nevertheless, although our results of contrast looks shallow in terms of contrast, the central stars are generally faint so our data can be as deep as those of the other previous studies in terms of limiting magnitude. Eventually, planets around YSOs are younger and brighter than those in the young moving groups, enabling us to constrain a few times higher than jovian mass.
On the other hand, our survey achieved the strongest constraints compared to other YSO imaging surveys (see Section \ref{sec: YSO Surveys}).

\begin{deluxetable*}{llcccccccc}
\tablewidth{0pt}

\tablecaption{SEEDS/YSO detection limits}
\tablehead{
\colhead{HD name} &
\colhead{Other name} &
\colhead{Magnitude (band) \tablenotemark{a}} &
\multicolumn{7}{c}{$\Delta$ mag (5$\sigma$ contrast)}
 \\
\colhead{} &
\colhead{} &
\colhead{} &
\colhead{0\farcs25} &
\colhead{0\farcs5} &
\colhead{0\farcs75} &
\colhead{1\farcs0} &
\colhead{1\farcs5} &
\colhead{2\farcs0} &
\colhead{3\farcs0}
}

\startdata
\dots&TYC 4496-780-1&7.76$\pm0.02$ ($H$)& 6.2 & 8.8 & 10.1 & 11.1 & \dots & 11.9 & 10.1 \\
21997&\dots&6.12$\pm0.03$ ($H$) & \dots & 7.2 & 9.1 & 10.4 & 12.0 & 12.7 & 12.9 \\
\dots&LkH$\alpha$ 330&7.92$\pm0.03$ ($H$)& \dots & 8.8 & 9.9 & 11.2 & 12.4 & 13.1 & 13.1 \\
\dots&IRAS 04028+2948&9.47$\pm0.03$ ($H$)& 3.9 & 5.5 & 6.3 & 7.1 & 7.6 & 7.6 & 6.1 \\
\dots&IRAS 04125+2902&9.76$\pm0.03$ ($H$)& \dots & 6.3 & 7.8 & 8.5 & 10.1 & 10.6 & 10.8 \\
\dots&LkCa 4, V1068 Tau& 8.52$\pm0.02$ ($H$)&\dots & 8.9 & 10.0 & 11.1 & 12.1 & 12.4 & 12.5 \\
\dots&Elias 1, V892 Tau&7.02$\pm0.03$ ($H$)& \dots & \dots & \dots & \dots & 9.7 & 10.6 & 11.5 \\
281934&BP Tau&8.22$\pm0.02$ ($H$)& \dots & 8.2 & 9.5 & 10.4 & 11.4 & 11.7 & 8.8 \\
\dots&V819 Tau&8.65$\pm0.02$ ($H$)& \dots & \dots & 8.3 & 9.4 & 10.7 & 11.3 & 11.0 \\
283571&RY Tau&6.13$\pm0.06$ ($H$)& \dots & 8.3 & 10.1 & 11.4 & 13.5 & 14.6 & 15.6 \\
284419&T Tau&6.24$\pm0.02$ ($H$)& \dots & \dots & \dots & \dots & \dots & \dots & 8.3 \\
\dots&LkCa 8, IP Tau&8.89$\pm0.02$ ($H$)& \dots & 7.8 & 8.7 & 10.1 & 11.0 & 11.1 & 10.8 \\
\dots&DG Tau&7.72$\pm0.03$ ($H$)& \dots & \dots & 8.5 & 9.5 & 10.9 & 11.5 & 8.2 \\
285846&UX Tau&7.96$\pm0.02$ ($H$)& \dots & \dots & \dots & \dots & \dots & \dots & 7.5 \\
\dots&HL Tau&7.41$\pm0.02$ ($K_{\rm s}$) & \dots & \dots & \dots & \dots & \dots & 7.3 & 4.6 \\
\dots&L1551-51, V1075 Tau&9.06$\pm0.02$ ($H$) & 8.8 & 9.9 & 10.9 & 11.2 & 11.4 & 11.4 & 9.7 \\
\dots&GG Tau(A)&7.82$\pm0.03$ ($H$)& \dots & \dots & \dots & \dots & \dots & \dots & 12.1 \\
\dots&L1551-55, V1076 Tau&9.46$\pm0.03$ ($H$)& 8.1 & 9.6 & 10.2 & 10.9 & 11.0 & 11.0 & 9.1 \\
\dots&DL Tau&7.96$\pm0.02$ ($K_{\rm s}$)& 6.9 & 8.9 & 9.6 & 10.1 & 10.7 & 10.8 & 9.2 \\
\dots&DM Tau&9.76$\pm0.02$ ($H$)& 5.0 & 7.8 & 8.4 & 9.4 & 10.3 & 10.6 & 11.0 \\
\dots&CI Tau&8.43$\pm0.04$ ($H$) &6.4 & 8.6 & 9.7 & 10.4 & 11.2 & 11.2 & 10.1 \\
\dots&DN Tau&8.34$\pm0.03$ ($H$)& 6.8 & 9.4 & 10.1 & 11.0 & 11.4 & 11.5 & 11.5 \\
\dots&LkCa 14, V1115 Tau&8.71$\pm0.03$ ($H$)& \dots & 8.6 & 9.7 & 10.8 & 11.8 & 12.1 & 12.2 \\
\dots&LkCa 15, V1079 Tau&8.60$\pm0.02$ ($H$) & 5.7 & 7.8 & 8.5 & 9.3 & 9.3 & 9.4 & 9.3 \\
\dots&LkH$\alpha$ 332/G1, V1000 Tau&8.40$\pm0.02$ ($H$) & \dots & \dots & 9.3 & 10.2 & 11.3 & 12.0 & 12.2 \\
\dots&GO Tau&9.78$\pm0.02$ ($H$)& 4.3 & 6.1 & 6.5 & 7.3 & 8.4 & 8.6 & 8.8 \\
\dots&GM Aur&8.60$\pm0.02$ ($H$)& 7.8 & 10.5 & 11.0 & 11.5 & 11.8 & 11.4 & 11.1 \\
282630&LkCa 19&8.32$\pm0.02$ ($H$)& 6.4 & 9.4 & 10.6 & 11.3 & 12.0 & 12.1 & 12.1 \\
31293&AB Aur&4.23$\pm0.02$ ($K_{\rm s}$)& 6.7 & 8.6 & 9.3 & 9.4 & 10.0 & 10.3 & 10.4 \\
282624&SU Aur&6.56$\pm0.02$ ($H$)& \dots & 11.1 & 12.2 & 13.0 & 14.1 & 14.3 & 12.0 \\
\dots&V397 Aur&8.32$\pm0.02$ ($H$)& 6.0 & 8.5 & 9.5 & 10.6 & 11.2 & 11.4 & 11.3 \\
\dots&V1207 Tau, RX J0458.7+2046&8.96$\pm0.02$ ($H$)& 6.7 & 9.0 & 10.3 & 10.9 & 11.8 & 11.9 & 11.9 \\
31648&MWC 480&6.26$\pm0.03$ ($H$)& 8.0 & 10.9 & 12.5 & 13.6 & 14.8 & 15.1 & 15.1 \\
34282&\dots&8.48$\pm0.03$ ($H$)& \dots & 8.5 & 9.7 & 10.8 & 12.0 & 12.3 & 12.4 \\
36112&MWC 758&6.56$\pm0.02$ ($H$)& \dots & 9.7 & 11.1 & 11.8 & 13.2 & 13.3 & 10.9 \\
36910&CQ Tau&7.06$\pm0.02$ ($H$)& \dots & \dots & \dots & 11.5 & 13.2 & 13.8 & 14.1 \\
290764&V1247 Ori&8.20$\pm0.25$ ($H$)& 7.6 & 10.0 & 11.4 & 12.2 & 13.2 & 13.5 & 13.7 \\
\dots&TW Hya&7.56$\pm0.04$ ($H$)& \dots & 7.2 & 8.9 & 9.6 & 10.6 & 10.8 & 9.6 \\
\dots&PDS 70&8.82$\pm0.04$ ($H$)& \dots & 6.9 & 8.4 & 9.8 & 10.7 & 11.0 & 10.9 \\
135344B&SAO 206462&6.59$\pm0.03$ ($H$)& \dots & 8.0 & 9.8 & 11.1 & 12.8 & 13.4 & 13.8 \\
139614&\dots&7.33$\pm0.04$ ($H$)& \dots & 7.4 & 9.5 & 10.7 & 12.9 & 13.7 & 14.2 \\
\dots&GQ Lup&7.70$\pm0.03$ ($H$)& \dots & 7.9 & 8.9 & 10.0 & 11.0 & 11.4 & 8.8 \\
141441&HIP 77545&8.00$\pm0.03$ ($H$)& 8.7 & 10.4 & 11.8 & 12.8 & 14.2 & 14.5 & 14.7 \\
142315&HIP 77911&6.67$\pm0.03$ ($H$)& \dots & 9.3 & 10.9 & 11.8 & 13.7 & 14.4 & 14.8 \\
\dots&IM Lup &8.09$\pm0.04$ ($H$)& 7.3 & 8.7 & 10.1 & 11.1 & 12.3 & 12.2 & 10.0 \\
142527&HIP 78092&5.72$\pm0.03$ ($H$)& \dots & 10.4 & 12.1 & 12.7 & 14.1 & 14.5 & 14.7 \\
\dots&RX J1603.9-2031A&8.77$\pm0.03$ ($H$)& \dots & 8.7 & 9.7 & 10.8 & 11.4 & 11.4 & 9.6 \\
\dots&RX J1604.3-2130A, USco J1604 &9.10$\pm0.02$ ($H$)& 7.8 & 9.7 & 10.9 & 11.8 & 12.4 & 12.4 & 11.9 \\
\dots&SZ 91&9.85$\pm0.02$ ($K_{\rm s}$)& \dots & 8.5 & 9.3 & 9.6 & 9.6 & 9.8 & 9.8 \\
144587&HIP 78996&7.36$\pm0.06$ ($H$)& \dots & 9.0 & 10.6 & 11.7 & 12.9 & 13.3 & 13.3 \\
\dots&V1094 Sco, RX J1608.6-3922&9.04$\pm0.02$ ($H$)& \dots & 8.3 & 9.6 & 10.7 & 11.7 & 11.8 & 12.0 \\
145655&HIP 79462&7.43$\pm0.07$ ($H$)& 7.6 & 9.1 & 11.1 & 12.8 & 14.1 & 14.3 & 14.4 \\
147137&HIP 80088&7.90$\pm0.04$ ($H$)& 7.7 & 9.7 & 11.1 & 11.9 & 12.8 & 13.0 & 13.2 \\
\dots&IRAS 16225-2607, V896 Sco&7.95$\pm0.06$ ($H$) & \dots & 8.0 & 11.9 & 12.3 & 11.3 & 11.5 & 10.4 \\
\dots&DoAr 25&8.40$\pm0.05$ ($H$)& \dots & 7.3 & 8.4 & 9.4 & 10.8 & 11.3 & 9.9 \\
148040&HIP 80535&7.31$\pm0.05$ ($H$)& \dots & 8.2 & 9.7 & 11.0 & 12.5 & 13.3 & 13.7 \\
\dots&DoAr 28&8.99$\pm0.02$ ($H$)& \dots & 8.4 & 9.8 & 10.6 & 11.2 & 11.4 & 8.5 \\
\dots&SR 21&7.51$\pm0.04$ ($H$)& \dots & 9.2 & 10.1 & 11.1 & 12.0 & 12.3 & 10.2 \\
\dots&IRAS 16245-2423, Oph IRS 48&8.82$\pm0.07$ ($H$)& \dots & 7.6 & 8.5 & 9.4 & 10.6 & 10.7 & 8.2 \\
\dots&DoAr 44, ROX 44&8.25$\pm0.06$ ($H$)& \dots & 8.6 & 9.2 & 10.3 & 11.3 & 11.4 & 9.1 \\
\dots&RX J1633.9-2442&9.36$\pm0.02$ ($H$)& \dots & 7.8 & 9.2 & 9.7 & 11.1 & 11.2 & 8.5 \\
169142&\dots&6.91$\pm0.04$ ($H$)& \dots & 9.8 & 11.1 & 12.2 & 13.3 & 13.6 & 12.9 \\
\dots&MWC 297&4.39$\pm0.21$ ($H$)& 6.4 & 8.7 & 10.5 & 11.6 & 13.6 & 14.6 & 15.0 \\
\dots&RX J1842.9-3532&8.71$\pm0.04$ ($H$)& \dots & 8.0 & 8.8 & 9.6 & 10.9 & 11.2 & 8.3 \\
\dots&RX J1852.3-3700&9.14$\pm0.02$ ($H$)& \dots & 9.5 & 10.6 & 11.1 & 11.8 & 11.9 & 11.7 \\
179218&\dots&6.65$\pm0.03$ ($H$) & \dots & \dots & \dots & \dots & \dots & \dots & 14.1 \\
200775&HIP 103763&5.47$\pm0.03$ ($H$)& \dots & 9.4 & 11.2 & 12.5 & 13.8 & 12.1 & 11.9 

\enddata
\tablenotetext{a}{$H$-band magnitudes taken from 2MASS \citep[][]{Cutri2003,Skrutskie2006}.}
\label{all contrast}
\end{deluxetable*}

\subsection{Luminosity-Mass Conversion and Detectable-Mass Limit}
\label{sec: Mass Estimation}
We reveal a detected companion's mass and how massive planets can be detected in our YSO observations.
Therefore, we need to convert our 5$\sigma$ contrast limit into the mass detection limit and the detected companion's luminosity into its mass.
The relationships between age, luminosity, and mass of giant planets and brown dwarfs have been theoretically modeled \citep[e.g.][]{Baraffe2003,Allard2011}. The relations should vary depending on how planets form from circumstellar disks \citep[][]{Marley2007}. 

We adopt the age-luminosity-mass relation of COND03 model \citep[][]{Baraffe2003}, one of hot-start models.   
Cold-start models have been also proposed to model the luminosity and temperature evolution of a giant planet, which depends on the planet formation scenarios \citep[e.g.,][]{Marley2007}.
Observational results disagree to a very cold start model \citep[][]{Marleau2014}.
As mentioned in Section \ref{sec: Introduction}, though the controversy of formation scenario has been unsettled, 
the mass estimations based on the evolution models should be uncertain particularly on young exoplanets, due to the large uncertainty of the initial conditions \citep[][]{Marleau2014}.
To calibrate these models requires comparing parameters between estimated by direct imaging and by other methods such as radial velocity or transit. However, there are no exoplanets detected by both of direct imaging and indirect methods so far.

The luminosity-mass conversions also require the age and distance of the target, so the stellar parameters of our observed YSOs should be known in our analysis.
As discussed in Section \ref{sec: SEEDS Targets and Selection Criteria}, most YSOs are located in star-forming regions, thus age and distance of YSOs can be approximately estimated from their group membership. Table \ref{star-forming-groups} shows typical age and distance of star-forming regions targeted in this study. Table \ref{each-parameter} shows the stellar parameters of each YSO. 
Note that we use the stellar parameters to estimate detection limits in Section \ref{sec: Results of Individual Companion Survey} and Section \ref{sec: Results and Discussion of Total Data}. The individually adopted parameters are usually consistent with those of the belonging groups. CQ Tau, IM Lup, Sz 91 have somewhat different distances from the those of Taurus and Lupus. V1094 Sco, HIP 79462, USco J1604, and LkH$\alpha$ 332/G1 have different ages from the typical value of Upper Sco and Taurus.
In this calculation, we used 10 Myr for only YSOs in Upper Sco instead of 11 Myr so as to avoid extrapolating the planet luminosity at 11 Myr from the luminosity models in 1--10 Myr.
For HIP 103763 we used 1 Myr instead of the adopted age because our adopted luminosity model does not publish the calculations for the objects with such a very young age.

\LongTables
\begin{deluxetable*}{llccccc}

\tablewidth{0pt}
\tablecaption{Adopted Stellar Parameters of SEEDS/YSO Targets}
\tablehead{
    \colhead{HD name} &
    \colhead{Other name} &
    \colhead{Sp type} &
    \colhead{Av} &
    \colhead{Age} &
    \colhead{Distance} &
    \colhead{Reference} \\
    \colhead{} &
    \colhead{} &
    \colhead{} &
    \colhead{} &
    \colhead{[Myr]} &
    \colhead{[pc]} &
    \colhead{}
    }

\startdata
\dots & TYC 4496-780-1 & G\tablenotemark{a} & \dots & 15 & 150 & 56  \\
21997 & \dots & A3 & \dots & 30 & 72 & 27,29,44,66,69  \\
\dots & LkH$\alpha$ 330 & G3 & 1.8 & 3 & 250 & 22,25,65,78  \\
\dots & IRAS 04028+2948 & A1 & \dots & 6 & 140 & 64 \\
\dots & IRAS 04125+2902 & M1.25 & 2.39 & 6 & 140 & 12,37,79 \\
\dots & LkCa 4, V1068 Tau & K7 & 1.21 & 2.5 & 140 & 1,13,18\\
\dots & Elias 1, V892 Tau & A6 & 5.93 & 2 & 140 & 1,18,63  \\
281934 & BP Tau & K7 & 0.49 & 2.2 & 140 & 1,18,45  \\
\dots & V819 Tau & K7 &1.35 &2.7 &140 &1,18,43,68 \\
283571 & RY Tau & K1 & 1.84 & 1.1 & 134 & 1,3,5 \\
284419& T Tau & K0 & 1.39 &1.8 &140 & 1,5,68  \\
\dots & LkCa 8, IP Tau & M0 & 0.24 & 4.0 & 140 & 1,11,40,80  \\
\dots& DG Tau& K6 &1.29 & 9.0 & 140 &12,68,79  \\
285846 & UX Tau & G8 & 0.21 & 1 & 152 & 1,5,6,45  \\
\dots& HL Tau &K7 & 7.4 & 0.9 & 140 &1,41,83  \\
\dots & L1551-51, V1075 Tau & K7 & 0.0 & 6 & 140 & 1  \\
\dots& GG Tau(A) &K7 &0.76 &1.5 &140 & 1,36  \\
\dots & L1551-55, V1076 Tau & K7 & 0.69 & 6 &140 & 1 \\
\dots& DL Tau & K7 & 1.21 & 1 & 140 & 3,12,17,43 \\ 
\dots & DM Tau & M1 & 0.00 & 3.6 & 140 & 1,3  \\
\dots& CI Tau & K7 & 1.77 & 1.7 &140 & 1,3,17  \\
\dots & DN Tau & M0 & 0.49 & 2.6 & 140 & 1,43  \\
\dots& LkCa 14, V1115 Tau & M0 & 0.00 & 8.9 & 140 & 1,43  \\
\dots & LkCa 15, V1079 Tau & K5 & 0.62 & 1.4 & 140 & 1,43\\
\dots& LkH$\alpha$ 332/G1, V1000 Tau & M1--M2.5 & 4.6 & 0.2 & 140 & 26,42,71 \\
\dots & GO Tau & M0 & 1.18 & 4.8 & 140 & 1,3  \\
\dots& GM Aur & K3--K5.5 & 1.2 & 7.2 & 140 & 3,20,84 \\
282630 & LkCa 19 & K0 & 0.00 & 6 & 140 & 1 \\  
31293& AB Aur & A0--A1 & 0.50 & 3 & 144 & 2,4,45 \\
282624& SU Aur & G2 & 0.90 & 2.2 & 140 & 1,3,45  \\
\dots& V397 Aur & K7 & 0.00 & 4.4 & 140 & 1,55,68,80,86  \\
\dots& V1207 Tau, RX J0458.7+2046 & K7 & 0.3 & 2.7 & 140 &52,81 \\
31648& MWC 480 & A5 & 0.096 & 6.7 & 146 & 2,31  \\
34282 & \dots & A3 & 0.16 & 6.4 & 350 & 2,9 \\
36112& MWC 758 & A5 & 0.0 & 3.5 & 279 & 47,49,50,66  \\
36910 & CQ Tau & F3 & 2.0 & 13 & 100 & 4,5  \\
290764& V1247 Ori & A0/A5/F2 & 0.64 & 7.4 & 385 & 62,89 \\
\dots & TW Hya & M2 & 0.0 & 8 & 54 & 57,58,66,77  \\
\dots& PDS 70 & K5 & 0.81 & 10 & 140 & 48,67,90,91 \\
135344B & SAO 206462 & F4 & 0.3 & 8 & 142 & 8,23,50,55,82  \\
139614& \dots & A7 & 0.09 & 7.0 & 140 & 8,82,85,92  \\
\dots &GQ Lup & K7--M0 & 1.0 & 3 & 150 & 15,72  \\
141441& HIP 77545 & A2--A3 & 1.25 & 11 & 119 & 14,66,93  \\
142315 & HIP 77911 & B9 & 0.34 & 11 & 148 & 34,66 \\
\dots& IM Lup & M0 & 0.5 & 1  & 190 & 19,94  \\
142527 & HIP 78092 & F6 & 0.37 & 2 & 140 & 2,85,87 \\
\dots& RX J1603.9-2031A & K5 & 0.7 & 11 & 145 &73\\
\dots & RX J1604.3-2130A, USco J1604 & K2 & 1.0 & 3.7 & 145 & 35,59,88  \\
\dots& SZ 91 & M0.5 & 2.0 & 5 & 200 &53,72,95  \\
144587 & HIP 78996 & A9 & 0.47 & 11 & 108 & 14,66,93  \\
\dots& V1094 Sco, RX J1608.6-3922 & K6 & 3.0 & 3 & 155 & 60,61  \\
145655 & HIP 79462 & G2 & 0.45 & 6 & 142 & 14,66,93  \\
147137 & HIP 80088 & A9 & 0.61 & 11 & 139 & 14,66,93  \\
\dots & IRAS 16225-2607, V896 Sco & K7 & 1.3 & 0.8 & 120 & 38,39  \\
\dots& DoAr 25 & K5 & 2.7 & 3.8 & 145 & 20,30,97,98  \\
148040 & HIP 80535& G0 & 0.00 & 8 & 120 & 14,66,74 \\
\dots& DoAr 28 & K5 & 2.3 & 5 & 139 &38,75 \\
\dots & SR 21 & G3 & 9.0 & 4.7 & 120 & 20,76  \\
\dots& IRAS 16245-2423, Oph IRS 48& A0 & 12.9 & 1.5 & 120 & 24,38  \\
\dots & ROXs 42B & M0 & 1.9 & 2.5 & 120 & 33,51  \\
\dots& DoAr 44, ROX 44 & K3 & 2.2 & 1.5 & 120 & 6,38,50 \\
\dots & RX J1633.9-2442 & K7 & 5.0 & 2.0 & 120 & 54,70  \\
169142 & \dots & A5 & 0.31 & 8 & 145 & 2,8,74  \\
\dots & MWC 297 & B1.5 & 8 & 2 & 250 &7,46,96  \\
\dots& RX J1842.9-3532 & K2 & 1.1 & 5 & 130 & 21,22,37,99  \\
\dots& RX J1852.3-3700 & K3 & 1.0 & 5 & 130 & 21,22,37,99 \\
179218 &\dots&B9 &0.77 &1 &240 &10,28,32  \\
200775 & HIP 103763 & B3 & 2.43 & 0.016 & 430 & 4,5,27,28

\enddata
\tablenotetext{a}{Inferred from its ${\rm T_{eff}}$ and the spectral type vs. ${\rm T_{eff}}$ relation of \citet[][]{Pecaut2013}}
\tablerefs{1. \cite{Hartmann1995} 2. \cite{Fukagawa2010} 3. \cite{Ricci2010} 4. \cite{Hernandez2004} 5. \cite{Bertout1999} 6. \cite{Espaillat2010} 7. \cite{Mora2001} 8. \cite{Dunkin1997a} 9. \cite{Merin2004} 10. \cite{Malfait1998} 11. \cite{Gullbring1998} 12. \cite{Andrews2013} 13. \cite{White_Ghez_2001_Taurus} 14. \cite{Pecaut2012} 15. \cite{McElwain2007} 16. \cite{Neuhauser2005} 17. \cite{Valenti1993} 18. \cite{Kucuk2010} 19. \cite{Pinte2008} 20. \cite{Andrews2009} 21. \cite{Silverstone2006} 22. \cite{Rigliaco2015} 23. \cite{Muller_2011_HD135344} 24. \cite{Marel2013} 25. \cite{Isella2013} 26. \cite{Herczeg2014} 27. \cite{Hipparcos_1997}  28. \cite{Alecian2013} 29. \cite{Moor2006} 30. \cite{Andrews2008} 31. \cite{Montesinos2009} 32. \cite{Panic2009b} 33. \cite{Bouvier1992} 34. \cite{Hernandez+05_HeAeBe} 35. \cite{Dahm2009} 36. \cite{White1999} 37. \cite{Hughes2010}  38. \cite{McClure2010} 39. \cite{Rojas2008} 40. \cite{Sestito2008} 41. \cite{Kraus2009} 42. \cite{Hartmann1998} 43. \cite{Takagi2014} 44. \cite{Torres_2008_moving_group} 45. \cite{Costigan2014} 46. \cite{Drew1997} 47. \cite{Benisty2015} 48. \cite{Hashimoto2012} 49. \cite{Meeus2012} 50. \cite{Andrews2011} 51. \cite{Currie2014a} 52. \cite{Wahhaj2010} 53. \cite{Tsukagoshi2014} 54. \cite{Orellana2012} 55. \cite{France2012}  56. \cite{Guillout2010} 57. \cite{Akiyama2015} 58. \cite{Herczeg2004} 59. \cite{Preibisch_99_USco} 60. \cite{Bustamante2015} 61. \cite{Joergens2001} 62. \cite{Kraus2013} 63. \cite{Hillenbrand1992} 64. \cite{Kenyon1990} 65. \cite{Brown2009} 66. \cite{Van2007} 67. \cite{Riaud2006} 68. \cite{Lopez2015} 69. \cite{Moor2013} 70. \cite{Cieza2010} 71. \cite{Rebull2010} 72. \cite{Hughes1994} 73. \cite{Carpenter2014} 74. \cite{Houk1982} 75. \cite{Rich2015} 76. \cite{Prato2003} 77. \cite{Debes_2013_TWHya} 78. \cite{Fernandez1995} 79. \cite{Luhman2010} 80. \cite{Bertout2007} 81. \cite{Wichmann_1996_Taurus} 82. \cite{vanBoekel_2005_HeBeStar} 83. \cite{White2004} 84. \cite{Espaillat2010} 85. \cite{Houk_1978_SptCatalog} 86. \cite{Leinert1993} 87. \cite{Mendigut_2014_HD142527}. 88. \cite{Kohler2000} 89. \cite{Caballero2008} 90. \cite{Gregorio2002} 91. \cite{Metchev2004} 92. \cite{Yudin1999} 93. \cite{Houk_1988_MKspec_type} 94. \cite{Wichmann_Hipparcos_1998} 95. \cite{Comeron2008} 96. \cite{Kaas2004} 97. \cite{Wilking2005} 98. \cite{Makarov2007} 99. \cite{Neuhauser2000} 
}
\label{each-parameter}
\end{deluxetable*}

Using the extinction law of \cite{Cardelli1989}, we first corrected the interstellar extinctions, which however only weakly influence most of our YSO targets ($<\sim$0.5 mag; see Table \ref{each-parameter}).
Adopted parameters of distance and age in Table \ref{each-parameter} are used o calculate mass in jovian mass and separation in AU. Figures \ref{mass_all_1} and \ref{mass_all_2} show 5$\sigma$ detection limit of mass with bright and faint YSOs as a function of AU. 
In this case, our observations could be sensitive to smaller exoplanets around the fainter YSOs than brighter YSOs, because the central stars are fainter.
This figure shows that we can set typical upper limit of 5--10 ${\rm M_J}$ at a few tens of AU. Our results can discuss exoplanets outside 100 AU and brown dwarfs in the solar-system scale. Because Orion Nebula Cluster, Serpens-Aquila and Perseus are located farther than the other star forming regions (see Tables \ref{star-forming-groups}, \ref{each-parameter}), it is unable to constrain planetary-mass companions within 100 AU in these star-forming regions. Some contrast curves are influenced by the bright companions (see Section \ref{sec: Contrast}), producing the biases in the contrast determinations.

\subsection{Results of Individual Companion Survey}
\label{sec: Results of Individual Companion Survey} 

\subsubsection{TYC 4496-780-1}
\label{sec: TYC 4496-780-1}
This isolated T Tauri star in front of the Cepheus complex has strong NIR and FIR excess and H$\alpha$ accretion signature \citep[][]{Guillout2010}.

SEEDS observed this system in 2012 July with qPDI+ADI mode. We detected a bright source at $\sim$1\farcs5.
\cite{Guillout2010} suggested that this system is a spectroscopic-binary system; we find that our detected object is possibly a stellar component of the pair.
Since we have no follow-up observations of this target and this detected source is very bright, we do not consider that this object is a planetary-mass companion candidate.

\subsubsection{HD 21997}
This A3 star is a member of the Columba moving group \citep{Moor2006,Moor2013}, which is as old as $\sim$30 Myr \citep{Torres_2008_moving_group}.  However, its SED shows a transitional-like disk rather than debris disk \citep[][]{Moor2013}. 
SEEDS therefore observed HD 21997 in the YSO category.

We reduced the data taken in 2013 January with sPDI+ADI and 0\farcs4 mask, detecting no companion candidates.

\subsubsection{LkH$\alpha$ 330}
LkH$\alpha$ 330 is a T Tauri star in the Perseus star forming region. This star has an SED of transitional disk, and a gap and asymmetric structure was observed in the disk \citep[e.g.][]{Brown2008,Brown2009,Isella2013}. 

SEEDS observed LkH$\alpha$ 330 three times in 2011 December using sPDI+ADI with 0\farcs4\ mask, 2014 October and 2015 January using qPDI+ADI mode without masks. Our data reductions for all data detected no companion candidate. We constrained the potentially existing exoplanet smaller than 60 ${\rm M_J}$ at 50 AU, 16.5 ${\rm M_J}$ at 100 AU.

\subsubsection{IRAS 04028+2948}
This Herbig Ae/Be star in the Taurus star forming region has an SED with the IR excess and concavity at MIR range \citep[][]{Kenyon1990, Rebull_2011_WISE_Taurus}.

We reduced the data observed in 2012 December with qPDI+ADI mode and detected no companion candidates in its FOV. We set a detection limit of 20 ${\rm M_J}$ at 50 AU.

\subsubsection{IRAS 04125+2902}
This T Tauri star is a 4\farcs0\ binary system in the Taurus star forming region and has a transitional disk \citep[][]{Kim2013,Espaillat2015}.

We reduced the data taken in 2013 November and in 2014 January with sPDI+ADI mode and confirmed this system to be a binary system. We did not detect any other companion candidates in its FOV.

\subsubsection{LkCa 4}
LkCa 4 is a T Tauri star in the Taurus star association and has an SED indicative of a class III object\citep[][]{Furlan2006,Howard2013}. The spectroscopic study derived Li abundance of this system \citep[][]{Sestito2008}.

SEEDS observations were carried out in 2012 November with qPDI+ADI, resulting in no detection of companion candidates in its FOV.

\subsubsection{Elias 1}
Elias 1 (= V892 Tau) is a Herbig Ae/Be star in a $\sim$4\arcsec\ binary system, which is a member of Taurus association, and that possesses a circumbinary disk \citep[][]{Monnier2008}, which is interesting to study the disk property. 

This object was observed in 2013 February with sPDI+ADI+0\farcs4 mask. We confirmed the stellar companion but did not detect any other companion candidates. 

\subsubsection{BP Tau} \label{sec: BP Tau}
This T Tauri star in the Taurus star forming region was observed by resolving the disk, in which CO appears to start to be depleted \citep[][]{Dutrey2003}.

SEEDS observations were carried out in 2012 September with qPDI+ADI mode. We detected a point source at $\sim$3\farcs1 near the edge of FOV.
In order to discuss proper motion of the companion candidate, we used HST/NICMOS and Subaru/CIAO data. Our CPM test showed that this source is likely a background star. We used the proper motions reported in \cite{Zacharias2012} ($\mu\alpha$=7.4$\pm$1.2, $\mu\delta$=-28.4$\pm$0.7 [mas/yr]). However, NICMOS data are saturated and CIAO data are masked, which means these data have large astrometric uncertainty. Judging whether this object is a companion or a background star requires follow-up observations.
We do not conclude that this object is a background star.

\subsubsection{V819 Tau}
V819 Tau is a class III object in the Taurus association and has an excess at 24 $\mu$m and FIR \citep[][]{Furlan2006,Luhman2009}, which may arise from a transitional disk.   

HiCIAO images taken in 2012 January with qPDI+ADI detected no companion candidates in its FOV.

\subsubsection{RY Tau}
RY Tau is a T Tauri star in the Taurus star forming region. 
\cite{Isella2010} resolved this disk and reported an inner gap, which suggests that a planet with a mass less than 5 ${\rm M_J}$ exists between 10 and 50 AU from the primary star.

HiCIAO images taken in 2011 January with sPDI+ADI+0\farcs4\ mask and in 2014 October with sPDI+ADI detected a faint source at 5\arcsec. We do not include this object as a companion candidate.

\subsubsection{T Tau}
T Tau was first-selected protostar as a new type of variable star \citep[][]{Joy1945}.
This YSO is a well-studied triple system in the Taurus-Auriga star forming region. T Tau is now called T Tau N and the companions are called T Tau Sa and T Tau Sb. 
These components of the system have IR excesses \citep[][]{Hogerheijde1997,Ratzka2009}.

SEEDS observation carried out in 2015 January had few rotation angle and we could not discuss exoplanets located within $2\farcs3$. 

\subsubsection{IP Tau}
This T Tauri star in the Taurus star forming region has a pre-transitional disk \citep[][]{Espaillat2011}.

SEEDS observation was carried out in 2011 December with qPDI+ADI mode. We reduced the data and detected no companion candidates. Calculation of detection limit excludes possible companions more massive than 14 ${\rm M_J}$ at 50 AU.

\subsubsection{DG Tau}
This T Tauri star is a $\sim$1\arcmin-separation binary system \citep[e.g.,][]{Rodmann2006} in the Taurus star forming region and has a resolved image of its circumstellar disk \citep{Isella2010}, accompanied by the radio jet \citep[][]{Lynch2013}.

We did not detect any point sources in the FOV of the qPDI+ADI data taken in 2015 January.

\subsubsection{UX Tau}
UX Tau A is a T Tauri binary in the Taurus star forming region. \cite{Espaillat2007} reported a transitional disk around UX Tau A. SEEDS reported that the polarization degrees largely vary over the disk, indicating a thin disk with dust grains \citep[][]{Tanii2012}.

We reduced the HiCIAO data taken in 2013 November with qPDI+ADI mode and confirmed the stellar companion at a separation of $\sim$2\farcs7. We did not detect any other companion candidates. For detection limit, we could not discuss the inner part of 400 AU, which were automatically masked by LOCI.

\subsubsection{HL Tau}
HL Tau is a T Tauri star associated with the Taurus star forming region. Previous studies using Subaru/CIAO \citep[][]{Tamura2000} has reported an asymmetric (C-shaped) feature and different color pattern on its disk \citep[][]{Murakawa2008}. Recently, multi ring feature in the mid-plane of the disk was reported by ALMA observations \citep[][]{ALMA2015}, with which \cite{Akiyama2016} discussed planet formation in this system. \cite{Testi2015}, using Large Binocular Telescope Interferometer mid-infrared camera (LBTI/LMIRcam), tried to find companions of HL Tau, but resulting in non detection and upper limit on exoplanets of $\sim$10--15 ${\rm M_J}$ at 70 AU.

SEEDS observations were conducted in 2015 January after ALMA press release. At the observations, the HL Tau's PSFs largely varied, possibly due to bad weather condition. In the final image made through our data reductions, there were residuals of stellar halo, preventing us from detecting any companion candidates.

\subsubsection{L1551-51}
L1551-51 is a T Tauri star in the Taurus association. Its SED classifies the star into class III object \citep[][]{Luhman2010,Howard2013}. \cite{Martin1994} derived Li abundances of this YSO from spectroscopic observations.

We reduced the data taken in 2012 November with qPDI+ADI and detected no companion candidates.

\subsubsection{GG Tau}
GG Tau is a T-Tauri star associated with the Taurus star forming region. This star is one of a quadraple system \citep[Aa, Ab, Ba, Bb][;]{White1999} and surrounded by the circumstellar disk \citep[e.g.,][]{Krist2005}. SEEDS reported the gap and asymmetric features in its disk \citep[][]{Itoh2014,Yang2016}. 

HiCIAO observed GG Tau twice, one of which were taken for short exposure time and LOCI cannot be applied. Thus we use only the same data as those presented by \cite{Itoh2014}, which were taken in 2011 September using sPDI+ADI mode and 0\farcs6 occulting mask. The occulting mask is large and this observation did not obtain the field rotation large enough to examine the inner region of the planetary system. As a result, the inner part within 300 AU is masked after the LOCI data reduction, detecting no companion candidate.

\subsubsection{L1551-55}
This T Tauri member of the Taurus star forming region is classified into class III \citep[][]{Luhman2010,Howard2013}. \cite{Magazzu1992} observed this YSO and estimated Li abundance.

SEEDS observation was carried out in 2012 November with qPDI+ADI mode. We did not detect any companion candidates in its FOV.

\subsubsection{DL Tau}
DL Tau is a T Tauri member of the Taurus association and has an infrared excess \citep[][]{Hartmann2005,Andrews2007a}. \cite{Andrews2007a} resolved the disk around DL Tau.

HiCIAO images observed in 2014 October with qPDI+ADI did not detect any companion candidates. 

\subsubsection{DM Tau}
This T Tauri star belongs to Taurus star forming region. Previous studies reported this system has a transitional disk \citep[][]{Calvet2005,Andrews2007a}. \cite{Andrews2011} reported a gap at $\sim$20 AU from the central star.

SEEDS observation was carried out in 2009 December with sPDI+ADI mode. Our data reduction detected no companion candidates.

\subsubsection{CI Tau}
CI Tau is a T Tauri star in the Taurus star forming region and has an infrared excess in its SED and its disk was resolved \citep[][]{Andrews2007a}. 

This system was observed in 2012 September with qPDI+ADI mode. We did not detect any companion candidates in its FOV.

\subsubsection{DN Tau}
DN Tau, a T Tauri member of the Taurus association, has an IR excess from MIR to FIR in its SED \citep[][]{Najita2007,Andrews2007a} and its disk was resolved by sub-mm observations \citep{Andrews2007a}.

We reduce HiCIAO data taken in 2009 December with sPDI+ADI, resulting in no detection of point sources.

\subsubsection{LkCa 14}
LkCa 14 is a T Tauri star in the Taurus association. This system is classified as class III object and has an IR exceess \citep[][]{Hartmann2005,Dent2013}, which may imply a transitional disk.

HiCIAO images taken in 2011 December with sPDI+ADI detected no companion candidates in the FOV.

\subsubsection{LkCa 15}
\label{sec: LkCa 15}
LkCa 15 is a T Tauri star in the constellation of Taurus and has a transitional disk \citep{Najita2007, Espaillat2007}, which has a gap structure reveled by various-wavelength imaging observations \citep[e.g.,][]{Pietu_2006_LkCa15,Thalmann2010,Thalmann2014,Thalmann2015}. Furthermore, the companion candidates have been reported and investigated by Keck/NIRC2 observation \citep[][]{Kraus2012}. The Large Binocular Telescope (LBT) confirmed the companion candidates and 
MagAO observation discovered the H$\alpha$ emission that implies mass accretion \citep[][]{Sallum2015}.

\cite{Thalmann2010} used the SEEDS data taken in 2009 December with sPDI+ADI mode. In addition to this observation, LkCa 15 was observed in 2010 January with ADI, in 2010 December with sPDI+ADI, in 2011 January with sPDI+ADI, and in 2013 November with qPDI+ADI mode. We analyzed all of these data and detected no point sources. The inner exoplanets detected by \cite{Sallum2015} are located in the strong self-subtraction region and LOCI automatically masked. 
In the first and fourth epoch, however, we recognized a similar pattern of signal in the final maps as Gemini/NIRI images \citep[][]{Thalmann2014}, though these patterns are low S/N ratio of $\sim$2--4. 
We think that these signals are protoplanetary disks but do not discuss the disk feature. We finally estimated upper limit of 5 ${\rm M_J}$ at 30 AU, 4.5 ${\rm M_J}$ at 50 AU, 3.5 ${\rm M_J}$ at 70 AU.

\subsubsection{LkH$\alpha$ 332 G1}
This T Tauri star is a $\sim$$0\farcs23$ binary system in the Taurus association \citep[][]{Leinert1993}. Previous studies reported accretion signatures \citep[e.g.][]{McCabe2006}. The SED has a IR excess which may represent a transitional disk \citep[][]{Hartmann2005}.

We observed this system in 2013 January with sPDI+ADI and $0\farcs4$ mask. The occulting mask prevented us from confirming the stellar companion. HiCIAO image detected a bright point source at the edge ($\sim$$10\arcsec$) of FOV and we think this object is a stellar companion or background star due to its wide separation.

\subsubsection{GO Tau}
This T Tauri star is a member of the Taurus star forming region. Its SED is appeared to be that of a transition disk \citep[][]{Najita2007}.  The disk was imaged at sub-millimeter wavelengths \citep[][]{Andrews2007a}.

SEEDS observed GO Tau in 2009 December with sPDI+ADI mode. We detect a point source at 4\farcs9, but did not re-observe this system. We exclude this object from exoplanet candidate because its separation is too large. We did not detect any other companion candidates.

\subsubsection{GM Aur}
GM Aur is a T-Tauri star of the Taurus association. HST/NICMOS resolved the protoplanetary disk \citep[][]{Schneider2003}. The SED of GM Aur represents a transitional disk \citep{Calvet2005,Najita2007,Espaillat2010}, whose cavity was detected by sub-mm observations \citep[e.g.][]{Hughes2009,Andrews2011}. Spectroscopy from FUV to NIR range has been executed \citep[][]{Ingleby2015} and theoretical simulation suggests a gap in the disk \citep[e.g.][]{Espaillat2010}. 

This object was observed in 2010 December with qPDI+ADI mode and 2011 December with sPDI+ADI mode. We reduced these data but detect no signal.
We calculated contrast of both data and show the better-contrast data on Table \ref{all contrast}. Finally, we estimated upper limit as 2.5 ${\rm M_J}$ at 50 AU and 1.5 ${\rm M_J}$ at 100 AU.

\subsubsection{LkCa 19}
This class III YSO is a T Tauri star in the Taurus star forming region \citep[][]{Hartmann2005,Howard2013}. This system has an IR excess at $\lambda = 24 {\rm \mu m}$ and can be thought to have a transitional disk \citep[][]{Hartmann2005,Luhman2010}.

We reduced the data observed in 2011 December with sPDI+ADI and detected a point source at $\sim$4\farcs4\ separation. 
We do not include this object among companion candidates due to its very wide separation.

\subsubsection{AB Aur}
AB Aur is a Herbig Ae/Be star in the Taurus association. This relatively-bright star has a protoplanetary disk resolved by various instruments \citep[e.g.,][]{Oppenheimer_2008_ABAur,Perrin_2009_ABAur}. The CIAO and HiCIAO observations revealed an asymmetric feature at $\sim$50--500 AU \citep[][]{Fukagawa2004,Hashimoto2011}. 

We reduced the data taken in 2009 December with sPDI+ADI mode that is different from \cite{Hashimoto2011}, who reported HiCIAO data taken in 2009 October with only PDI mode. 
We did not detect any point sources. We then calculated upper limit of the mass of potentially existing exoplanets to be 13 ${\rm M_{J}}$ at 100 AU.

\subsubsection{SU Aur}
SU Aur is a T Tauri star associated with the Taurus association. This system had been reported to have an IR excess \citep[e.g.][]{Hartmann2005} and a nebulosity \citep[][]{Nakajima1995}.
SEEDS revealed the asymmetric and tail structures in this system \citep[][]{DeLeon2015}.

SU Aur was observed in 2014 January and in 2014 October with qPDI+ADI mode. We reduced these data and did not detect any companion candidate. Detection limit is estimated to be 10 ${\rm M_J}$ at 50 AU.

\subsubsection{V397 Aur}
This class III YSO in the Taurus star forming region has an IR excess \citep[][]{Hartmann2005,Furlan2006}, which is indicative of a transitional disk.

SEEDS observation was carried out in 2011 December with sPDI+ADI mode and detected a point source at $\sim$$6\arcsec$. We do not include it in companion candidates because of very wide separation.

\subsubsection{RX J0458.7+2046}
RX J0458.7+2046 (V1207 Tau) is a T Tauri star in the Taurus star forming region. \cite{Wichmann2000} estimated Li abundance of this system and \cite{Wahhaj2010} reported this system has a diskless SED.

HiCIAO observation was conducted in 2012 November with qPDI+ADI mode and in 2015 December with sPDI+ADI. We detected a point-like source at the edge of FOV. The follow-up observation in 2015 showed the source was false positive and we did not detect any companion candidates within 400 AU.

\subsubsection{MWC 480}
MWC 480 is a relatively-bright, Herbig Ae/Be star in the Taurus association and has a full disk so far imaged at multiple wavelengths \citep[][]{Pietu_2006_LkCa15,Kusakabe2012,Grady2010_MWC480}, with the observations of CO emissions from the disk \citep{Akiyama2013}.

We reduced the data observed in 2010 January with sPDI+ADI mode and detected 2 sources at $\sim$$4\arcsec$ and 2 sources at $\sim$$6\arcsec$, which are not included in planetary-mass companion candidates due to their large separations.

\subsubsection{HD 34282}
HD 34282 is a Herbig Ae/Be star associated with the Orion Nebula Cluster. It has a large disk \citep{Merin2004} with the CO emissions \citep{Dent2005}, whose image was previously resolved \citep{Pietu_2003_HD34282}. A recent study, published after our SEEDS observations for HD 34282, suggested that the disk is a transition disk \citep{Khalafinejad_2016_HD34282}.

SEEDS observation was carried out in 2011 December with sPDI+ADI and $0\farcs4$ mask. We detected no companions within its FOV. 

\subsubsection{MWC 758}
MWC 758 is a Herbig Ae/Be star in the Taurus star forming region. This system has been known to have a transitional disk \citep[][]{Isella2010b} and CO emission in its disk \citep[][]{Dent2005}. \cite{Grady2013} and \cite{Benisty2015} reported the obvious asymmetric features and spiral arms, which can be caused by giant planets. \cite{Dong2015} discussed the potential exoplanets that can make the asymmetric and arm features.

SEEDS observed MWC 758 in 2011 December with sPDI+ADI plus 0\farcs4 mask and qPDI+ADI, and 2013 October with qPDI+ADI mode. We analyzed these data and detected a point source at $\sim$2\farcs5, which was concluded as a background star by \cite{Grady2013}. We did not detect any other companion candidates and finally exclude the possibility that there is an exoplanet with a mass larger than 16 ${\rm M_J}$ outside 150 AU.

\subsubsection{CQ Tau}
This Herbig Ae star in the Taurus association has a disk in which CO emission was reported \citep[][]{Dent2005}. By investigating carbon chemistry of the disk, \cite{Chapillon2010} suggested that CQ Tau's disk is likely a transitional disk.  Previous studies discussed radial profile of the disk \citep[e.g.][]{Doucet2006,Trotta2013}.

SEEDS observation carried out in 2012 January with qPDI+ADI, in 2014 October with sPDI+ADI+$0\farcs3$ mask and 2015 December with sPDI+ADI. We detected a point source at $\sim$$2\farcs2$ and our CPM test revealed that this object is likely a background star. We adopted the proper motions reported in \cite{Zacharias2012} ($\mu\alpha=2.2\pm0,6$, $\mu\delta=-25.5\pm0.7$ [mas/yr]).

\subsubsection{V1247 Ori}
Previous spectral classifications of V1247 Ori range from F2 to A0 \citep[e.g.][]{Vieira2003,Kraus2013}. V1247 Ori is a member of the Orion star forming region and shows a dip around $\sim$15 $\mu$m in the SED, suggesting the presence of a transitional disk \citep[][]{Caballero2010,Kraus2013}. \cite{Kraus2013} discovered an asymmetric feature in the disk using the Very Large Telescope Interferometer, the Keck Interferometer, Keck-II, Gemini South, and IRTF.

SEEDS observations were conducted three times in 2013 November with sPDI+ADI plus $0\farcs3$ occulting mask, in 2014 January with qPDI+ADI and 2015 January with ADI mode. We reduced these data and did not detect any companion candidate. Since this object is located at 385 pc, thus our constraints cannot discuss exoplanets within 100 AU; our observations can constrain the mass of potentially existing objects of 33 ${\rm M_J}$ at 100 AU and 12 ${\rm M_J}$ at 200 AU. 

\subsubsection{TW Hya}
TW Hya is the nearest system from the Earth among our whole targets. This T Tauri star is in the TW Hydrae moving group \citep[e.g.][]{Webb1999,Zuckerman2001,Brandt2014b}, which is located about 50 pc and is about 10 Myr years old. The declination is very low ($\delta < -30^\circ$), making it relatively difficult to observe this moving group in high airmass. 
TW Hya has a transitional disk with active accretion reported by its SED \citep[e.g.][]{Calvet2002,Goto2012,Menu2014} and its disk has been resolved in various wavelengths \citep[][]{Krist2000,Weinberger2002,Qi2004}. 
Detailed information of selection criteria is described in \cite{Akiyama2015}.
SEEDS also observed this system and revealed multi-gap feature at 20 and 80 AU \citep[][]{Akiyama2015}. 

We analyzed the same data as reported in \cite{Akiyama2015}, which were taken in 2011 March and 2013 January with qPDI+ADI mode. We did not detect any signal and set constraints on detectable mass of exoplanet as 16 and 3 ${\rm M_J}$ at the inner and outer gaps.
\cite{Akiyama2015} analyzed the HiCIAO polarized intensity (PI) images and estimated a mass of potential planet, which creates a gap structure, to be lower than 0.7 ${\rm M_J}$ according to the theoretical calculations of \cite{Jang2012}. Our results are consistent with the predictions of \cite{Jang2012} but cannot set stronger constraints.

\subsubsection{PDS 70}
\label{sec: PDS 70}
PDS 70 is a T Tauri star in the Centaurus group. The SED of this system implies a transitional disk \citep[][]{Metchev2004,Riaud2006} and \cite{Riaud2006} resolved a scattered light disk around PDS 70 (detailed information is described in \cite{Hashimoto2012}).

SEEDS observed this object in 2012 February with qPDI+ADI mode and detected a gap feature \citep{Hashimoto2012}. This gap is so wide as $\sim$70 AU that \cite{Hashimoto2012} discuss the possibility of multi planets system. We reduced the same data as mentioned above, detecting a point source at $\sim$2\farcs2.
\cite{Hashimoto2012} concluded this source is a background star by combining their results of observations and those in \cite{Riaud2006}. Thus we did not detect any companion candidate. We estimated the detection limit of 16 ${\rm M_J}$ at 70 AU.

\subsubsection{SAO 206462}
SAO 206462 is a T Tauri star, sometimes called as Herbig F star, and a $\sim$20\arcsec visual binary system \citep[HD 135354;][]{Augereau2001}. The secondary star is outside the FOV of SEEDS observation. This object is in the constellation of Lupus. SAO 206462 has been predicted to have a gap from its SED feature \citep[][]{Brown2007}, which was later confirmed by sub-millimeter imaging \citep{Brown2009}.

\cite{Muto2012} revealed spiral arms and asymmetric feature using the SEEDS data, and also discussed potential planets by examining the disk's geometry. The Submillimeter Array (SMA) and ALMA observations also revealed the inner gap \citep[][]{Brown2009,Marel2016}. \cite{Dong2015} discussed the potential exoplanets that can make the asymmetric features around SAO 206462, as well as MWC 758. 

SEEDS observed this system in 2011 May with sPDI+ADI and $0\farcs4$ occulting mask. After the LOCI data reduction, we detected two point sources. Our identified sources are exterior to 400 AU from SAO 206462. Furthermore, our observations are unable to detect a companion candidate in the gap of disk, due to strong self-subtraction of LOCI-ADI reductions.
Finally we set upper limit. \cite{Muto2012} predicted the positions of two potential exoplanets, one of which however is in the software mask of LOCI and cannot be detected even if this planet exactly exists. At 0\farcs9, $\sim$130 AU where another potential planet could be located, HiCIAO detection limit excludes the probability of planets larger than 3 ${\rm M_J}$. This constraint agrees to the theoretical prediction of mass of the planet \citep[][]{Muto2012}.

\subsubsection{HD 139614}
This star is a Herbig Ae/Be star associated with the Lupus-Ophiuchus complex, with the dust emissions in the SED \citep[][]{Meeus_1998_HD139614,Matter2014} that can be interpreted as a pre-transitional disk  \citep[][]{Matter2014}.

We reduced the data acquired in 2014 January with sPDI+ADI and 0\farcs4 mask and detected 2 point sources at $\sim$3\farcs5 separations. We also detected relatively low SN ($\sim$4--4.5) objects within 1\arcsec and need follow-up observation. Therefore we did not detect convincing companion candidates within 400 AU.

\subsubsection{GQ Lup}
\label{sec: GQ Lup}
GQ Lup is a T-Tauri star in the Lupus star forming region and has the infrared excess with MIR concavity \citep[][]{Dai2010}. In this system, a companion (GQ Lup b) has been discovered by direct imaging \citep[][]{Neuhauser2005}. The mass estimate of GQ Lup b ranges from 1 to 60 ${\rm M_J}$ \citep[e.g.][]{Neuhauser2005,Seifahrt2007a,Lavigne2009}. 

\begin{figure}
    \centering
    \includegraphics[scale=0.25]{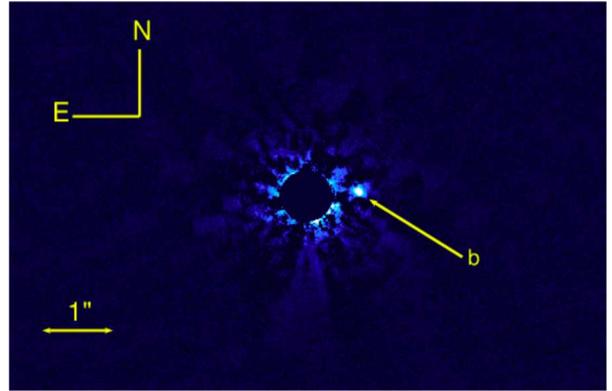}
    \caption{Subaru/HiCIAO $H$-band image of GQ Lup b taken in 2013 May. The central star is masked. North is up and east is left.}
    \label{GQ Lup b}
\end{figure}

SEEDS observed this planetary system in 2013 May with both qPDI+ADI mode and sPDI+ADI mode plus 0\farcs4 occulting mask.
We confirmed the companion (see Figure \ref{GQ Lup b} and Table \ref{detected companions}).  Our qPDI+ADI data reductions derived the separation and position angles of GQ Lup b to be ($\rho$, $\theta$) = (0\farcs723$\pm$0\farcs012, 277\fdg38$\pm$1\fdg40).
Our results did not deviate from the measurements in \cite{Neuhauser2008} but had larger errors of position angle because the seeing was bad and central star of science frames are masked or strongly saturated. 
\cite{Neuhauser2008} found that the separation between the companion and central star decreases by $\sim$2--3 mas/yr due to its orbital motion \citep[Figure 2 (a) of ][]{Neuhauser2008}. The extrapolation of the separation decrease calculated by \citet{Neuhauser2008} is consistent with our data, which can help to constrain the companion's orbital motion.

Our photometric estimation resulted in deriving a mass of GQ Lup b to be $\sim$15--20 ${\rm M_J}$.
We also detected one more point source at 6\arcsec but this object has been concluded as a background star \citep[e.g.][]{Neuhauser2008}. We also excluded the possibility of exoplanets more massive than 16 ${\rm M_J}$ at 70 AU.

\hspace{0.5cm}

\subsubsection{HIP 77545}
This Herbig Ae/Be star is a member of Upper Sco. \cite{Carpenter2009} detected an infrared excess at 24 $\mu$m but did not detect the excesses at 8 and 16 $\mu$m.

SEEDS observations were in 2012 May with sPDI+ADI and 2014 January with ADI. We detected a point source at $3\farcs8$ and two additional sources at the wider separations in the second data. On the other hand, the first data detected no point sources with the significant SN ratios and need more follow-up observations. Given that this system is located about 120 pc, we do not include these sources among companion candidates.

\subsubsection{HIP 77911}
HIP 77911 is a Herbig Ae/Be star in the Upper Sco association. Previous study has reported a debris disk \citep[][]{Carpenter2009,Mathews2013}.

This system was observed in 2012 May with sPDI+ADI and 2014 January with ADI mode. We reduced the data and detect some point sources at separations larger than $4\arcsec$ ($\sim$600 AU). We do not count them as companion candidates.

\subsubsection{IM Lup}
IM Lup is a T Tauri star in the Lupus association. The IM Lup's protoplanetary disk has been imaged from optical to radio wavelengths \citep[][]{Pinte2008,Panic2009a}. The surface density of the disk greatly changes at $\sim$ 400 AU \citep[][]{Panic2009a}.

IM Lup was observed in 2014 June with qPDI+ADI mode. After the data reduction, we detected two companions at 2\arcsec and 2\farcs5. \cite{Mawet2012} has reported the point source closer to the central star as a background star by calculating its proper motion with the VLT/NACO and HST/NICMOS data, but the more distant one has not been reported, requiring more follow-up observations for the clarification of this object.

\subsubsection{HD 142527}
HD 142527 is a relatively-bright, Herbig Ae/Be star in the constellation of the Lupus association. Some studies have reported the gap and asymmetric structure of the disk in this system \citep[e.g.][]{Fukagawa_2006_HD142527,Verhoeff2011,Fukagawa2013,Avenhaus2014} The possibility of companion existences was examined by \citep[][]{Biller2012}. 

SEEDS observed HD 142527 in 2011 May with sPDI+ADI combined with 0\farcs4 mask and in 2012 April with sPDI+ADI mode. We detected 2 high-SNR point sources and one marginally-detected source in each data set, which are located at distances larger than 550 AU. Therefore we exclude them from the list of companion candidates. Furthermore, we estimated the upper limit of detectable companion mass of 20 ${\rm M_J}$ at 50 AU.

\subsubsection{RX J1603.9-2031A}
This T Tauri star is one of a $\sim$4\arcsec\ binary system in the Upper Sco association \citep{Kohler_2000_USco} and shows an infrared excess \citep[][]{Carpenter2009,Mathews2013}.

SEEDS images taken in 2012 July with qPDI+ADI detected no companion candidates and set upper limit of potential exoplanet to be 5 ${\rm M_J}$ at 50 AU from the central star.

\subsubsection{RX J1604.3-2130A}
RX J1604.3-2130A (USco J1604) is a T Tauri star of the Upper Sco association and has an SED indicative of transition disk \citep[][]{Dahm2009}. This transition disk shows a gap feature revealed by SMA and Subaru (SEEDS) \citep[][]{Mathews2012,Mayama2012}.
\cite{Mayama2012} reported the presence of an arc-like structure, inside RX J1604's ring, at the west side of the star. This structure is now suspected to be an artifact.

We reduced HiCIAO data observed in 2012 April with qPDI+ADI mode, which have been used by \cite{Mayama2012}.
We did not detect any companion candidate and set detection limit of the mass of potentially existing exoplanet as 5.5 ${\rm M_J}$ at 50 AU.

\subsubsection{SZ 91}
This T Tauri star is a member of the Lupus molecular cloud  \citep[][]{Hughes1994} with a transitional disk \citep[e.g.][]{Merin2008,Romero2012}. \cite{Tsukagoshi2014} also used the HiCIAO data and spatially resolve the inner ($r\sim$65 AU) and outer ($r\sim$170 AU) disks.

We used the same data as reported by \cite{Tsukagoshi2014}, which were taken in 2012 May with sPDI+ADI mode. As a result, we detected two point sources at $\sim$$4\farcs2$ and $\sim$$9\arcsec$. We do not count them in companion candidates and set upper limit of 6.5 ${\rm M_J}$ at 70 AU.

\subsubsection{HIP 78996}
This Herbig Ae/Be star associated with the Upper Sco association has a debris disk \citep[][]{Carpenter2009,Mathews2013}.

SEEDS observations were carried out in 2011 May with sPDI+ADI+$0\farcs4$ mask and in 2013 May with sPDI+ADI. We detected a point source at $3\farcs5$ from the primary star. Our CPM test indicates that this object is likely a background star. For the proper motion test, we used the proper motions reported in \cite{Zacharias2012} ($\mu\alpha=-11.8\pm0.9$, $\mu\delta=-23.2\pm0.9$ [mas/yr]).

\subsubsection{V1094 Sco}
This T Tauri star in the Lupus association has a full disk \citep[][]{Tsukagoshi2011,Bustamante2015}.

We reduced the data taken in 2011 May with sPDI+ADI+$0\farcs4$ mask and detected 5 point sources at $5\farcs5$--$9\arcsec$ ($\sim$850--1400 AU). The outer radius of the disk was estimated to be 320 AU \citep{Tsukagoshi2011}; we therefore do not discuss possibility of companion for these sources.

\subsubsection{HIP 79462}
\label{sec: HIP 79462}
HIP 79462 is a T Tauri star in the Upper Sco association. This system has a debris disk \citep[][]{Carpenter2009, Dahm2012}. 

SEEDS observed HIP~79462 three times with only ADI mode in 2012 April, 2014 April and 2015 April. We detected a companion, HIP 79462B (see Figure 4 and Table 6) at ($\rho$, $\theta$) = (0\farcs648$\pm$0\farcs013, 3\fdg35$\pm$1\fdg07); these values are derived from 2012 April data. The $H$-band contrast is $4.2\pm0.3$ magnitudes; this gives an absolute $H$-band magnitude of $5.9\pm0.3$ assuming a distance of 142 pc and an $H$-band magnitude of $7.43\pm0.07$ for the primary \citep[][]{Cutri2003,Skrutskie2006}. We have no measurements of the companion's color. The COND03 and Solar-metallicity BT-Settl models at an age of 10 Myr \citep[][]{Baraffe2003,Allard2011} give an effective temperature of $\sim$3200 K and a mass of 0.2 $M_\odot$, well above the minimum mass for hydrogen ignition. \cite{Pecaut2013} list a spectral type of M4 at $T_{\rm eff} = 3200\ {\rm K}$.

\begin{figure}
    \centering
    \includegraphics[scale=0.25]{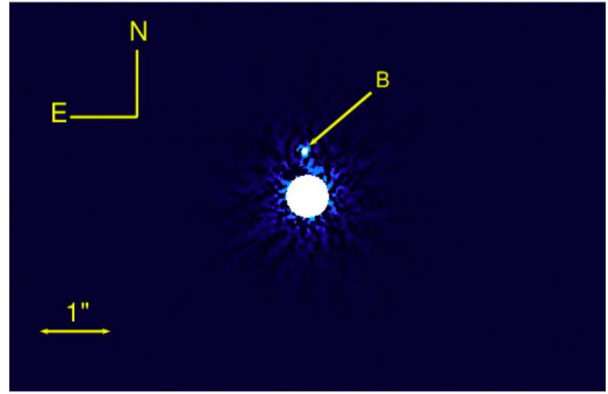}
    \caption{Subaru/HiCIAO $H$-band image of a companion around HIP 79462 taken in 2014 April. The central star is masked. North is up and east is left.}
    \label{HIP 79462}
\end{figure}
\hspace{0.5cm}

\subsubsection{HIP 80088}
HIP 80088 is a Herbig Ae/Be star in the Upper Sco association. Its SED indicates the presence of a debris disk \citep[][]{Carpenter2009,Mathews2013}.

This object was observed in 2012 July twice, using ADI-only and sPDI+ADI mode. We detected 2 point sources at $\sim$8\arcsec\ and 10\arcsec\ but do not include them among companion candidates due to their wide separations.

\subsubsection{IRAS 16225-2607}
\label{sec: IRAS 16225}
This T Tauri star is a member of the Ophiuchus star forming region. The previous study reported an SED that appears to be a transitional disk's SED \citep[][]{Furlan2009}. \cite{Furlan2009} did not report a stellar companion in this system, but \cite{Elliott2015} and \cite{Andsell2016} reported a stellar companion at $\sim$0\farcs9.

SEEDS observation was in 2012 November with qPDI+ADI mode. We detected the same source at $\sim$1\farcs0\ separation as \cite{Elliott2015} reported. Photometric result showed this object is more massive than brown dwarf ($\Delta H \sim1$ mag). Another point source was detected, which is located at 1\farcs5\ from the central star and near the bright stellar companion, and was therefore possibly affected by the companion, needing the follow-up observations.

\subsubsection{DoAr 25}
DoAr 25 is a T-Tauri star in the constellation of Ophiuchus and its SED indicates less evolved version of transitional disk \citep[][]{Andrews2008}. 

We reduced the 2 data taken in 2012 May with qPDI+ADI mode and 2014 January with ADI mode. We do not detect any signal in the first data, but detected a companion candidate in the second data. This object is separated by $4\farcs5$ from the central star, and we thus assume that this is not a substellar companion. Our calculated contrast exclude the possibility of exoplanets more massive than 13 ${\rm M_J}$ at 70 AU.

\subsubsection{HD 148040}
This T Tauri star is a member of the Upper Sco association. This system has a 24 $\mu$m excess but luck a 70 $\mu$m excess \citep[][]{Chen2005,Chen2011}.

HiCIAO images were taken in 2012 May with sPDI+ADI and in 2015 April with ADI mode.
We detected a point source at $\sim$$3\farcs9$ but the CPM test revealed this object to be a background star. We used the proper motions reported in \cite{Zacharias2012} ($\mu\alpha=-11.4\pm1.4$, $\mu\delta=-22.0\pm1.0$ [mas/yr]).

\subsubsection{DoAr 28}
DoAr 28 is a T Tauri star associated to the $\rho$ Ophiuchus association. This system has a transitional disk with an asymmetric and hook feature \citep{McClure2010, Rich2015}.  
A companion candidate discovered at $\sim$1\arcsec was concluded to be a background star using the SEEDS data \citep[][]{Rich2015}.

We reduced the same data as \cite{Rich2015} taken in 2012 May and 2014 January with qPDI+ADI mode. We detected the companion candidate and conducted the CPM test, confirming this object to be a background star.

\subsubsection{SR 21}
SR 21 is a T Tauri star of the Ophiuchus star forming region and is classified as one of a $\sim$6\farcs4 binary system \citep[][]{Prato2003}. This disk has a transitional disk feature and a cavity identified by the SED data and sub-mm observations \citep[][]{Furlan2009,Andrews2009,Brown2009,Perez2014,Marel2016}. Nevertheless, \cite{Follette2013} reported no cavity in the NIR scattered-light image.

For SEEDS observation, SR 21 was observed in 2011 May with qPDI+ADI mode. We reduced this data set and detected a feature like a point source, which has a SN ratio of 4.5. However, there are the remnants of spiders in the final image and this source is detected near these remnants. Therefore we assume this signal is likely effected by the spiders. We also ruled out possibility of existing exoplanet smaller than 11 ${\rm M_J}$ at 70 AU.

\subsubsection{Oph IRS 48}
Oph IRS 48 is a Herbig Ae/Be star associated with the Ophiuchus association. The disk of its system shows a gap feature \citep[][]{Brown2012}, which possibly originates from a companion \citep[][]{Marel2013,VanderMarel2014}. 

We reduced the SEEDS data taken in 2013 May with qPDI+ADI mode and did not detect any signal. We then set upper limit of 30 ${\rm M_J}$ at 50 AU.

\subsubsection{ROXs 42B}
\label{sec: ROXs 42B}

ROXs 42B is a T Tauri star in the $\rho$ Ophiuchus star forming region. This system has a companion ROXs 42B b and a companion candidate ``c'' \citep[][]{Currie2014a}. We follow the expression of ``c'' in \cite{Currie2014a} because \cite{Kraus2014} and \cite{Currie2014b} independently argued that this object is likely a background star.

\begin{figure}
    \centering
    \includegraphics[scale=0.25]{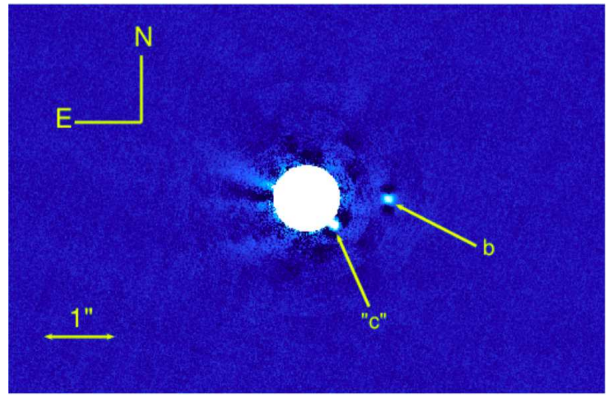}
    \caption{Subaru/HiCIAO $Y$-band image of a companion around ROXs 42B taken in 2014 June. The central star is masked. North is up and east is left.}
    \label{ROXs42B}
\end{figure}

SEEDS observed ROXs 42B in 2014 June with ADI mode. We reduced the data and confirmed the companion b and ``c'' (see Figure \ref{ROXs42B} and Table \ref{detected companions}). They are located at ($\rho$, $\theta$) = (1\farcs14$\pm$0\farcs004, 269\fdg7$\pm$0\fdg17) and (0\farcs55$\pm$0\farcs006, 228\fdg7$\pm$0\fdg64), respectively. Our results will help discuss the orbital motion of the companion. However, there were no reference frames in this observation and these errors may be underestimated. This follow-up observation after \cite{Currie2014a,Currie2014b} was conducted in $Y$-band ($\sim$0.97--1.07 ${\rm \mu m}$), which is different from typical HiCIAO imaging band. Therefore we use this data as confirmation of the companion and do not discuss the detection limit.

\hspace{0.5cm}

\subsubsection{ROX 44}
\label{sec: ROX 44}
ROX 44 is a T Tauri star associated with the Ophiuchus star forming region. This system has a (pre-)transitional disk \citep[][]{Andrews2009, Espaillat2010}.

We analyzed SEEDS data taken in 2011 May with qPDI+ADI and 2012 May with sPDI+ADI mode. We detected a point source at $\sim$2\farcs3 in the first-epoch data.
In the second-epoch data, this companion candidate is located at the inner region where self-subtraction is significant. We conducted the classical ADI data reduction \citep[][]{Marois2006} with the second data to avoid self-subtraction, but did not detect the candidate. 
We also have a trouble in making the fits header at the first-epoch data, in which the angular offset was mistakenly recorded. Thus we had to assume the offset from the other parameters in the header.
We used HST/WFC3 data taken in 2014 October to check proper motion. We subtracted the $180^\circ$ rotated image from the original image and confirmed a point source at a similar position. 
The PSF of central star is heavily saturated, making the centroid measurements uncertain.
Because of the large uncertainties and the trouble in our fits header, we still include this object among companion candidates and we need the more follow-up observations. We finally set lower limit of detectable mass of potential companions to be 3 ${\rm M_J}$ at 70 AU.

\subsubsection{RX J1633.9-2442}
This T Tauri star is a member of the Ophiuchus molecular cloud and has a transitional disk \citep[][]{Cieza2010, Cieza2012, Orellana2012}.

SEEDS observed this object in 2014 April with qPDI+ADI mode and detected no companion candidates in its FOV. We then estimated detection limit to be 9 ${\rm M_J}$ at 50 AU.

\subsubsection{HD 169142}
HD 169142 is a Herbig Ae/Be type, relatively bright star isolated from star forming regions \citep[][]{Dent2006}. Thus, we adopt the stellar parameters individually estimated for HD 161492 itself (see Table \ref{each-parameter}). The HD 169142's disk has a gap \citep[][]{Grady2007,Quanz2013b} and an asymmetric feature \citep[][]{Quanz2013b}, and a companion candidate \citep[][]{Reggiani2014}. 

\begin{figure}
    \centering
    \includegraphics[scale=0.25]{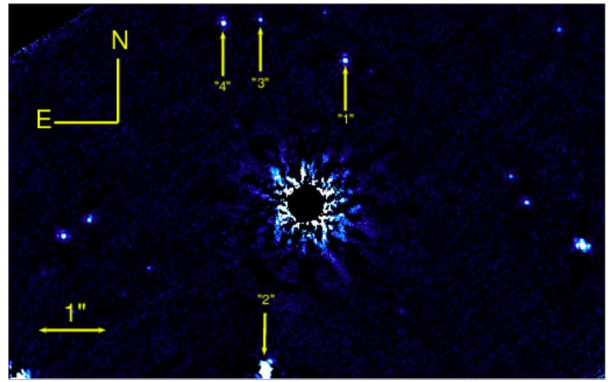}
    \caption{Subaru/HiCIAO $H$-band image of companion candidates around HD 169142 taken in 2011 May. The central star is masked. North is up and east is left. Sources which are not pointed by arrows are outside 400 AU or have low SN. The common proper motions of the No. 1, 2, 3, and 4 sources are tested (see Figure. \ref{cpmtest HD169142}). }
    \label{HD169142}
\end{figure}

\begin{figure}
    \centering
    \includegraphics[scale=0.48]{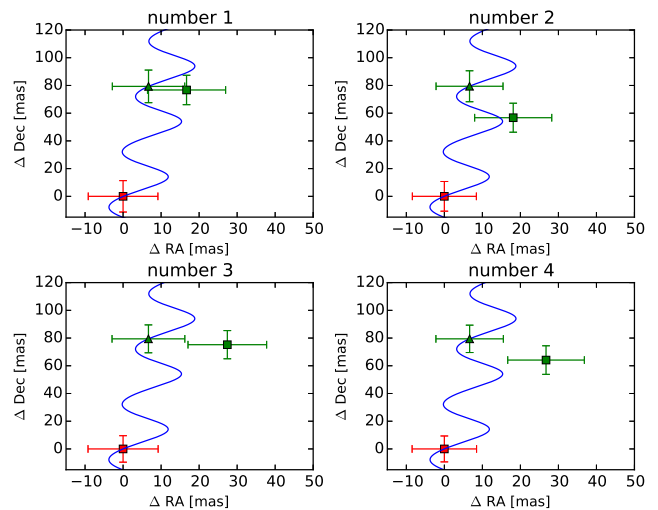}
    \caption{The vertical and horizontal axes are the $\Delta\alpha$ and $\Delta\delta$ of the companion candidates relative to their first-epoch positions.The number represents the object 1,2,3, and 4 labeled in Figure \ref{HD169142}, respectively. The red square plots correspond to the observed positions at the first epoch (2011 May 23), while the green square plots correspond to the positions observed at the second epoch (2013 May 19). If the companion candidates are background stars, their $\Delta\alpha$ and $\Delta\delta$ changes along with the blue curves, on which the green triangle plots indicate the predicted positions of background stars at the second epoch.}
    \label{cpmtest HD169142}
\end{figure}

\begin{figure*}
\centering
\centering\includegraphics[width=0.75\textwidth]{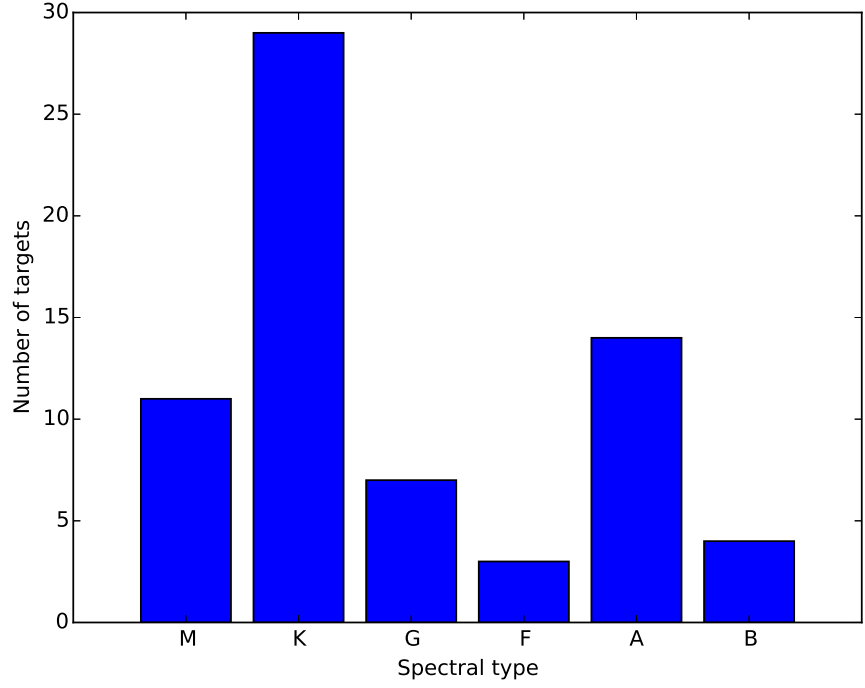}
\caption{Distribution of spectral types of YSOs that we targeted in this work.}
\label{splist}
\end{figure*}

\begin{figure*}
\centering
\centering\includegraphics[width=0.75\textwidth]{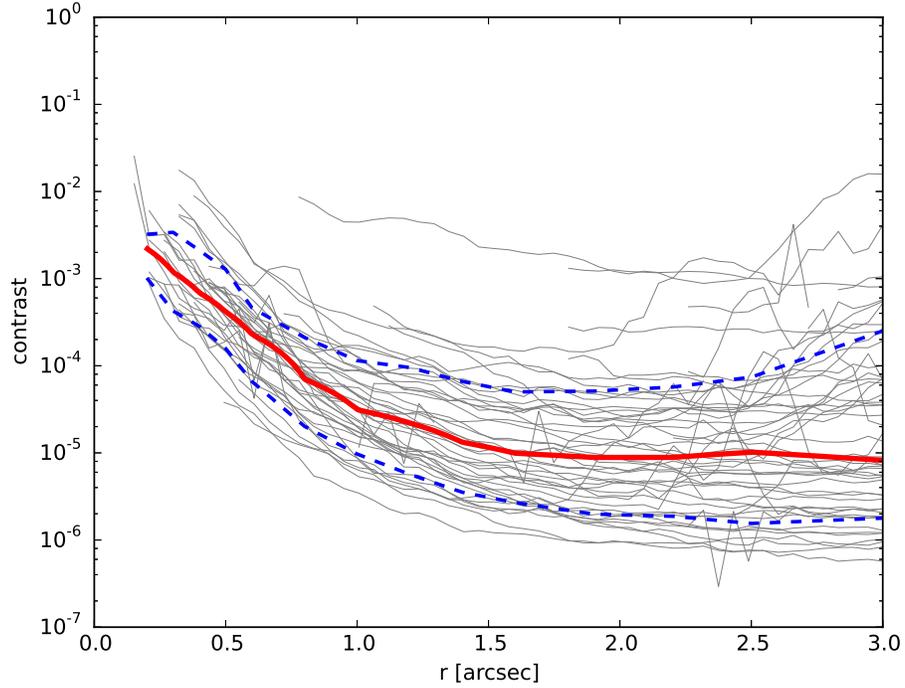}
\caption{Radial profile of contrast curves that are brighter than 8 mag in $H$ or $K_{\rm{s}}$ band (gray); the red line represents their median contrast curve and the two blue dashed lines represent 1 $\sigma$ range.
The horizontal axis represents a separation in arcsec and the vertical axis is a contrast to the central star.
We do not show contrast curves further than 3\arcsec\ because of the qPDI limitation.}
\label{contrast_all_1}
\end{figure*}

\begin{figure*}
\centering
\centering\includegraphics[width=0.75\textwidth]{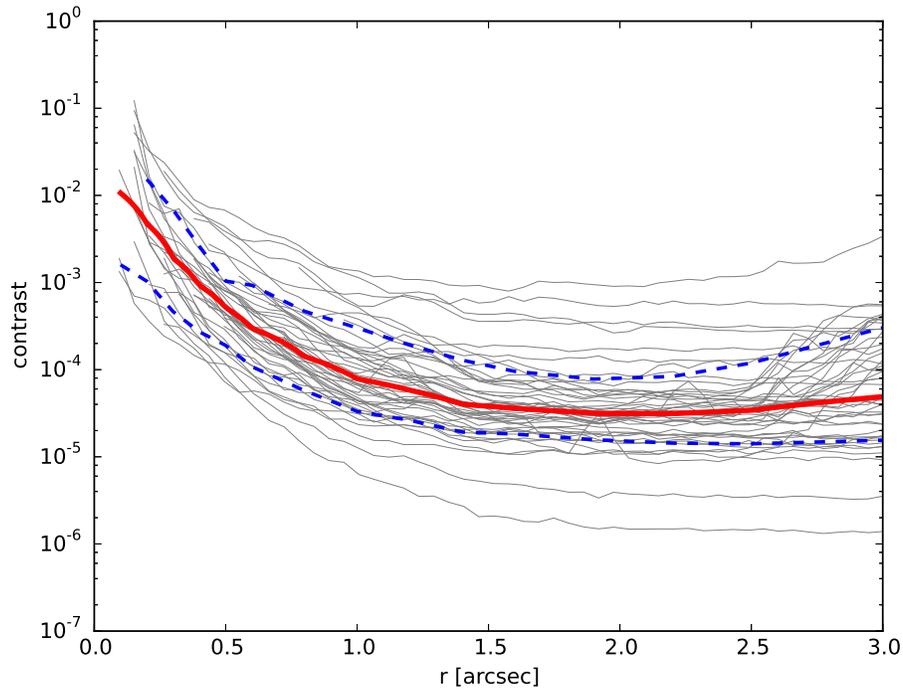}
\caption{Same as Figure \ref{contrast_all_1} except that our targets' brightness magnitudes at the observed band are not brighter than 8 mag.}
\label{contrast_all_2}
\end{figure*}

\begin{figure*}
\centering\includegraphics[width=0.75\textwidth]{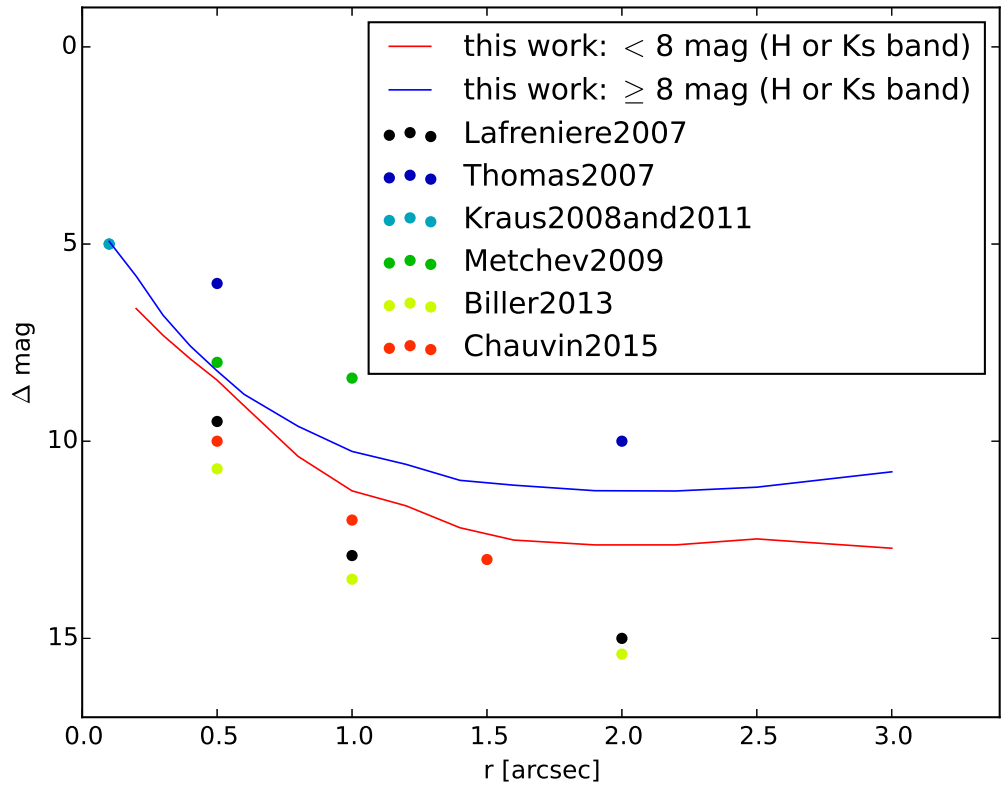}
\caption{Comparing typical contrast limits between this work and the other YSO and moving group surveys. The studies labeled in this Figure are described in Section \ref{sec: Comparison with Previous Surveys}. We roughly investigated plots by checking contrast curves in each study.} 
\label{comparison}
\end{figure*}

\begin{figure*}
\centering\includegraphics[width=0.75\textwidth]{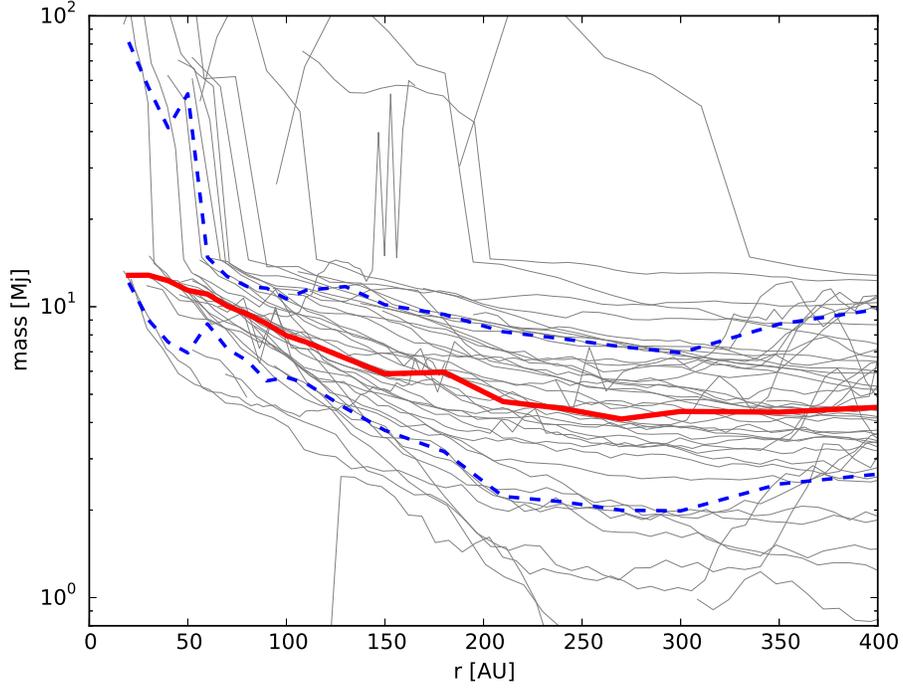}
\caption{Detection limits of each YSO in unit of mass. These gray lines are converted from the results in Figure \ref{contrast_all_1} using COND03 model \citep[][]{Baraffe2003}. The median value of the detection limits is represented by the red line and the 1$\sigma$ range is showed by the blue dashed lines. The vertical and horizontal axes represent the mass of 5$\sigma$ detection limit and the separation in AU, respectively.}
\label{mass_all_1}
\end{figure*}

\begin{figure*}
\centering\includegraphics[width=0.75\textwidth]{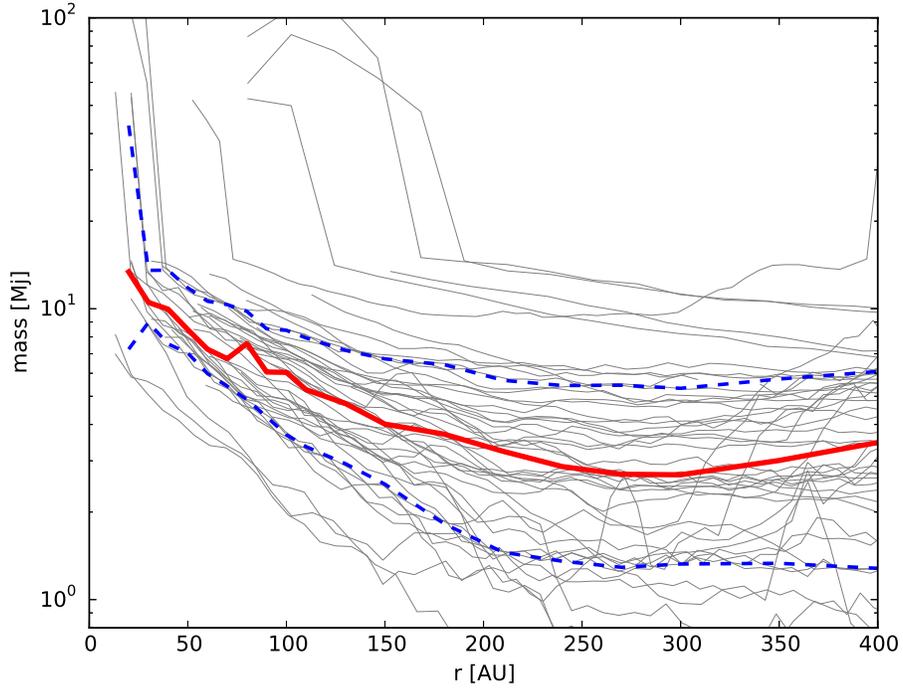}
\caption{Same as Figure \ref{mass_all_1} except that lines are converted from those in Figure \ref{contrast_all_2}.}
\label{mass_all_2}
\end{figure*}

SEEDS observations were conducted in 2011 May and 2013 May with sPDI+ADI plus a 0\farcs4 mask. The PDI data reduction detected a gap and an asymmetric feature in its disk \citep[][]{Momose2015}. On the other hand, SEEDS images cannot discuss the reported companion candidate which is located within the occulting mask. Other than this companion candidate, we detected a lot of companion candidates (see Figure \ref{HD169142}). HD 169142 is located within the Sagittarius constellation and there are a lot of background stars. We then select signals within 400 AU from all detected sources, which resulted in 6 companion candidates in 2011 May. However, only 4 candidates of these objects were detected in the follow-up observation in 2013 May. Then we executed CPM test for the 4 companion candidates; their separations and position angles are listed in Table \ref{bcgs}. Figure \ref{cpmtest HD169142} shows the results of the CPM test. The first observation did not take position reference frame and we may underestimate errors. We consider them to be background stars because they move so fast and similarly. 
The difference can be explained by their proper motion and parallax.
We assumed proper motion of ($\mu_\alpha$, $\mu_\delta$) = (-3.5$\pm$1.6, -39.9$\pm$1.3) [mas/yr] \citep[][]{Zacharias2012}.
The two objects remaining without the proper motion tests will be investigated with the next follow-up observations. For detection limit, we constrained the mass of potential companions to 16.5 ${\rm M_J}$ at 50 AU.

\subsubsection{MWC 297}
This Herbig Ae/Be star belonging to the Serpens-Aquila star forming region has a (pre-)transitional disk \citep[][]{Malbet2007,Rumble2015}. We adopted the distance and age of this YSO from the typical value of the star forming region \citep[see Table \ref{each-parameter};][]{Kaas2004}.

HiCIAO images taken in 2012 July with sPDI+ADI mode detected 2 companion candidates at the separations of $\sim$3\arcsec and $\sim$3\farcs4 ($>$750 AU). We do not include them in exoplanet candidates due to their large separations. We achieved contrast of $10^{-5}$ at 1\arcsec but we were not able to attain the detection limits good enough to detect a planetary-mass companion within 400 AU due to the bright central star.

\subsubsection{RX J1842.9-3532}
RX J1842.9-3532 is a T Tauri star associated with the CrA molecular cloud complex and has a transitional disk \citep[][]{Hughes2010}.

We used the HiCIAO data observed in 2012 September with qPDI+ADI and detected no companion candidates in its FOV. We then constrain the mass of possible exoplanets to 11.5 ${\rm M_J}$ at 50 AU.

\subsubsection{RX J1852.3-3700}
RX J1852.3-3700 is a T Tauri star in CrA and has a transitional disk \citep[][]{Hughes2010}. 

SEEDS observation was conducted in 2011 September with sPDI+ADI mode. We detected 6 companion candidates; only one of them is within 3\arcsec ($\sim$400 AU) and we exclude the other 5 point sources from companion candidates. This companion candidate also needs follow-up observation to confirm if it is a companion or background object. For detection limit, we set the mass limit of substellar companions to be 6.5 ${\rm M_J}$ at 50 AU.

\subsubsection{HD 179218}
HD 179218 is an isolated Herbig Ae/Be star and located at 240 pc from the Sun. The SED has an IR excess that seems to have a disk with an inner polar cavity \citep[][]{Elia2004,VanDerPlas2008}. \cite{Liu2007} resolved its disk and suggested an inner hole of the disk.

The HiCIAO data of this object were taken in 2012 September with sPDI+ADI and 0\farcs4 mask in SEEDS. Our data reduction revealed more than 10 point sources at the radial distances larger than 550 AU. We assume these point sources as background stars or stellar companions. Inner part of $2\farcs3$ ($\sim$550 AU) is masked due to the insufficient rotation angle, and thus we cannot derive the detection limit within 400 AU.

\subsubsection{HIP 103763}
HIP 103763 is a triple system in the Cepheus complex. 
The primary and secondary stars constitute the spectroscopic binary and the separation between each component is 15 mas \citep[e.g.][]{Millan-Gabet2001,Okamoto2009}. In addition, the third component is separated by $\sim$6\arcsec\ from the primary \citep[][]{Li1994}.
The primary star has a flare disk that extends to $\sim$1000 AU \citep[][]{Okamoto2009}.

HiCIAO observation was conducted in 2011 September with sPDI+ADI+$0\farcs4$ mask. We confirmed the bright star at $\sim$$3\arcsec$, which can be a background object or the third companion.

\section{Results and Discussion of Total Data}
\label{sec: Results and Discussion of Total Data}

\subsection{Companion Candidates}
Tables \ref{detected companions} and \ref{bcgs} list all identified point sources within $\sim$400 AU. Other companion candidates are described in Section \ref{sec: Results of Individual Companion Survey}.
GQ Lup, ROXs 42B, HD 169142 and LkCa 15 have been reported to have the candidates of substellar companion including brown dwarf. The observational conditions are explained in Section \ref{sec: Results of Individual Companion Survey}. We detected only GQ Lup b and ROXs 42B b (see Table \ref{detected companions}), and because objects around HD 169142 and LkCa 15 are located in the inner region where LOCI automatically covers the mask. Furthermore, we did not detect any companion candidates around YSOs whose disks have intriguing features suggesting planet formation such as SAO 206462, MWC 758, and HD 142527.

\begin{deluxetable*}{llcccc}
\tablecaption{Detected Companions in This Study}
\tablehead{
\colhead{HD name} &
\colhead{Other name} &
\colhead{$\Delta$ H (mag)} &
\colhead{Separation} &
\colhead{Position Angle}&
\colhead{Date (HST) and Method}}
\startdata
\dots&GQ Lup&6.7$\pm$0.3&0\farcs723$\pm0$\farcs012&277\fdg38$\pm$1\fdg40&2013 May 17, qPDI+ADI\\
&&&0\farcs723$\pm$0\farcs034&275\fdg66$\pm$1\fdg53&2013 May 19, sPDI+ADI\\\hline
145655 & HIP 79462 & 4.2$\pm$0.3 & 0\farcs648 $\pm$0.013&3\fdg35$\pm$1\fdg07&2012 Apr 11, ADI\\
&&&0\farcs676$\pm$0\farcs010&4\fdg26$\pm$0\fdg86&2014 Apr 22, ADI\\
&&&0\farcs666$\pm$0\farcs011&4\fdg87$\pm$0\fdg78&2015 Apr 29, ADI\\ \hline
\dots&ROXs 42B&\dots\tablenotemark{a}&1\farcs14$\pm$0\farcs004&269\fdg7$\pm$0\fdg17&2014 June 8, ADI
\enddata
\tablenotetext{a}{This system was observed in $Y$-band.}
\label{detected companions}
\end{deluxetable*}

\begin{deluxetable*}{llccc}
\tablecaption{Detected Background Stars in This Study}
\tablehead{
\colhead{HD name} &
\colhead{Other name} &
\colhead{Separation} &
\colhead{Position Angle} &
\colhead{Date (HST)}}
\startdata
36910&CQ Tau&2\farcs16$\pm$0\farcs02 &53\fdg1$\pm$0\fdg5&2012 Jan\\\hline
144587&HIP 78996&3\farcs49$\pm$0\farcs01&219\fdg9$\pm$0\fdg2&2013 May\\\hline
148040&HIP 80535&3\farcs83$\pm$0\farcs01&175\fdg8$\pm$0.1&2015 Apr\\\hline
169142 &\dots& 2\farcs21$\pm$0\farcs01 & 344\fdg8$\pm$0\fdg3 &2011 May\\
&& 2\farcs56$\pm$0\farcs01 & 166\fdg3$\pm$0\fdg2 &\\
&& 2\farcs82$\pm$0\farcs01 & 14\fdg1$\pm$0\fdg2& \\
&& 2\farcs96$\pm$0\farcs01 & 24\fdg7$\pm$0\fdg2&
\enddata
\label{bcgs}
\end{deluxetable*}

\subsection{Preliminary Statistical Discussion Assuming only 2 Companions} \label{sec:preliminary_frequency}

We confirmed 2 convincing low-mass companions at $\sim$100--150 AU out of 68 YSOs. The frequency can be simply computed to be 2/68, corresponding to a probability of $\sim$2.9\% for the substellar companions with masses between 1 and 70 ${\rm M_J}$ and projected separations between 50 and 400 AU. \par  
We note that the estimate of the detection frequency of substellar companions is very preliminary, since our calculation ignored several factors needed in the statistical discussion, such as the differences of detection sensitivities and the orbital parameters of companions. To statistically improve these problems in the analysis of the frequency of substellar companions, the Bayesian analysis should be useful \citep[e.g.,][]{Brandt2014a,Biller2013,Lafreniere2007b}. We will describe the statistical framework of Bayesian analysis in the forthcoming paper. 

\subsection{Comparison with Previous Surveys}
\label{sec: Comparison with Previous Surveys}
\subsubsection{Surveys for Nearby Young Stars}
\label{sec: Young Moving Group Surveys}
Apart from SEEDS, there are some planet-finding programs that targeted more nearby and relatively older targets, such as stars in young moving groups, than those in star forming regions.
The Gemini Deep Planet Survey \citep[][]{Lafreniere2007b} observed 85 nearby stars with ages $<$ 6700 Myr and distances $<$ 100 pc. This survey obtained 5$\sigma$ $H$-band detection limits of 9.5 mag at 0\farcs5, 12.9 at 1\arcsec, 15.0 at 2\arcsec and 16.5 at 5\arcsec, which are deep enough to detect companions more massive than 2 ${\rm M_J}$ with projected separations ranging from 40 to 200 AU.
The Gemini NICI Planet Finding survey \citep[][]{Biller2013} carried out high contrast imaging of 80 nearby stars with ages younger than several hundreds of Myr. They achieved 5$\sigma$ completeness detection limit of 10.7, 13.5, and 15.4 mag at 0\farcs5, 1\arcsec, and 2\arcsec, which are estimated from 95\% threshold of detection limits reported in \cite{Biller2013} by using equation (1) in \cite{Brandt2014a}.
The VLT/Naco Large Programe \citep[][]{Chauvin2015} observed solar-type targets with distances smaller than 100 pc and ages younger than 200 Myr, leading to obtain the typical contrasts of $\sim$10, 12, and 13 mag at 0\farcs5, 1\arcsec, and 1\farcs5, respectively. Other direct imaging surveys for stars older than YSOs ($\sim$10--5000 Myr) are also summarized in \cite{Chauvin2015}.
Our observations targeted YSOs, not nearby young stars, and our results are useful for comparisons with those older than YSOs.
\par

The roughly-estimated probability of substellar companions derived in Section \ref{sec:preliminary_frequency} can be compared with the probability of previous surveys around older stars. 
\citet[][]{Brandt2014a} performed a statistical analysis for the high-contrast data taken by the SEEDS and other teams. The statistical analysis for the sample of $\sim$250 nearby stars including a lot of young stars in moving groups suggests that 1.0--3.1\% (68\% confidence) or 0.92--11\% (95\% confidence) of stars host substellar companions with masses of 5--70 ${\rm M_J}$ between 10 and 100 AU \citep[][]{Brandt2014a}.
Our tentative results look apparently in agreement with this conclusion.
The main difference between our study and \cite{Brandt2014a} is the age range and separation range; this difference can be useful for discussing a formation rate of substellar companions and a possibility of their orbital evolutions.
If the statistical analysis for YSOs will result in disagreement with those of older stars such as \cite{Brandt2014a}, that disagreement would arise from a difference of conditions in the planet formation or evolution mechanism.

\subsubsection{YSO Surveys}
\label{sec: YSO Surveys}
\cite{Thomas2007} observed 72 Herbig Ae/Be stars with Gemini/Altair-NIRI and VLT/NACO. In \cite{Thomas2007}, some companions including brown dwarfs were detected but gas giant exoplanets were not detected. They achieved $K$-band detection contrasts of $10^{-2.4}$ at 0\farcs5 and $10^{-4}$ at farther than 2\arcsec; our obtained contrasts are much higher than theirs. The number of our targets is comparable with \cite{Thomas2007}.
\cite{Metchev2009} conducted Paloma/Keck adaptive optics survey around 266 F5-K5 stars between 3 Myr and 3 Gyr and between 10 pc and 190 pc, among which $\sim$30 targets are YSOs. They statistically derived the frequency of brown dwarfs (0.012--0.072 ${\rm M_\odot}$) to be $3.2^{+3.1}_{-2.7}$\% (2$\sigma$ confidence).
\cite{Fukagawa2010} observed 16 Herbig Ae/Be stars with Subaru/CIAO and obtained 2$\sigma$ $H$-band contrast of $\sim$9.5 mag at 2\arcsec\ and $\sim$12 mag beyond 4\arcsec.
SEEDS provides the contrast limits better and more YSO (including T Tauri stars) observations than in \cite{Fukagawa2010}. 
\cite{Kraus2008,Kraus2011} have conducted AO-aided high-resolution imaging surveys at Taurus and Upper Sco with Keck/NIRC2 and Paloma/PHARO observatories to reveal brown dwarf frequency.
Typical detetion limit of these studies are above the deutrium-burning limit except that the detection limits for some YSOs can constrain an exoplanet with a mass of a few ${\rm M_J}$.

As mentioned above, detecting planetary-mass object near YSOs was very difficult due to inadequate contrasts at small separations.
Although targets of \cite{Kraus2008,Kraus2011} overlap with ours, this study is the first large scale and systematic exploration for exoplanets around YSOs. Although our detection limits are still ought to be updated, their youth enables to constrain the mass of detectable exoplanet up to a few of jovian mass.

\section{Summary}
\label{sec: Summary}
SEEDS project completed main survey of exoplanets and disks around more than 400 targets. Out of these targets, we reduce and analyze the 99 SEEDS/YSO data sets for exoplanet exploration.
Our exploration provides the first large-scale and systematical analysis that can achieve high contrast enough to detect young exoplanets, and uses the largest sample among YSO exoplanet survey to the present. YSOs often have protoplanetary disks where planets are being formed. SEEDS project adopts the PDI technique combined with the ADI technique that are effective to detect both the disks and exoplanets around YSOs.  

We discovered 15 new point sources within 400 AU around 68 YSOs; among those, 7 are identified to be background stars, one is identified to be a stellar companion, and one is either a stellar companion or background star. 
We will aim at follow-up observations of remaining 6 companion candidates to conduct their CPM tests.
We also confirmed 2 companions which are previously reported; GQ Lup b and ROXs 42B b. LKCa 15 b, c and HD169142 b are not detected in this research.
We estimated 5$\sigma$ detection limits of contrast as a function of separation angle, which are transformed into detection limits in mass based on the properties of the observed YSOs (see Table \ref{each-parameter}) and the COND03 models. As a result, we typically achieved contrasts of $\sim$$10^{-3.5}$ at 0\farcs5, $10^{-4}$--$10^{-5}$ at 1\arcsec, and $10^{-4.5}$--$10^{-6}$ beyond 2\arcsec. 
We can also determine typical upper limits of potentially existing exoplanets to be $\sim$10 ${\rm M_J}$ at 0\farcs5 and $\sim$6 ${\rm M_J}$ at 1\arcsec.

Since previous studies calculated older-exoplanet frequency than our work \citep[e.g.][]{Brandt2014a,Chauvin2015}, our study will help compare the results of imaging surveys that targeted different ages.

Furthermore, we expect the next `extreme' adaptive optics system such as Gemini/GPI \citep[][]{Macintosh2006}, VLT/SPHERE \citep[][]{Beuzit2006}, and SCExAO \citep[][]{Guyon2010} to reveal a relationship between giant exoplanets and disk feature exploring the gap region in protoplanetary disks.

\acknowledgments

The authors thank David Lafreni$\grave{e}$re for generously providing the source code for the LOCI algorithm.
The authors would like to thank the anonymous referees for their constructive comments and suggestions to improve the quality of the paper.
This research is based on data collected at the Subaru Telescope, which is operated by the National Astronomical Observatories of Japan.  
Based in part on data collected at Subaru telescope and obtained from the SMOKA, which is operated by the Astronomy Data Center, National Astronomical Observatory of Japan.
We also acknowledge the SDPS project for our use of the CIAO archival data.
Some/all of the data presented in this paper were obtained from the Mikulski Archive for Space Telescopes (MAST). STScI is operated by the Association of Universities for Research in Astronomy, Inc., under NASA contract NAS5-26555. Support for MAST for non-HST data is provided by the NASA Office of Space Science via grant NNX09AF08G and by other grants and contracts.
Data analysis were carried out on common use data analysis computer system at the Astronomy Data Center, ADC, of the National Astronomical Observatory of Japan. 
This research has made use of NASA's Astrophysics Data System Bibliographic Services.
This research has made use of the SIMBAD database, operated at CDS, Strasbourg, France. 
This research has made use of the VizieR catalogue access tool, CDS, Strasbourg, France. The original description of the VizieR service was published in A{\&}AS 143, 23.
This publication makes use of data products from the Two Micron All Sky Survey, which is a joint project of the University of Massachusetts and the Infrared Processing and Analysis Center/California Institute of Technology, funded by the National Aeronautics and Space Administration and the National Science Foundation.

MK was supported by Japan Society for Promotion of Science (JSPS) Fellowship for Research and this work was partially supported by the Grant-in-Aid for JSPS Fellows (Grant Number 25-8826).
JC is supported by the U.S. National Science Foundation under Award No. 1009203.
MT is partly supported by the JSPS Grant-in-Aid (Nos. 15H02063 and 22000005).

The authors wish to recognize and acknowledge the very significant cultural role and reverence that the summit of Mauna Kea has always had within the indigenous Hawaiian community. We are most fortunate to have the opportunity to conduct observations from this mountain.

\clearpage
\bibliographystyle{apj}                                                               
\bibliography{library}                                                                

\end{document}